\documentclass[a4paper]{article}

\usepackage{graphicx}
\usepackage{subfigure}
\usepackage{floatflt}
\usepackage{afterpage}
\usepackage{array,longtable}
\def\NIM{Nucl. Instr. Meth.}

\def\NIM{Nucl. Instr. Meth.}

\newcommand{\BW}{\mathrm{BW}}
\newcommand{\GS}{\BW^{\mathrm{GS}}}
\newcommand{\mrho}{\mathrm{M}_\rho}
\newcommand{\momg}{\mathrm{M}_\omega}

\begin{document}

\setcounter{tocdepth}{3}

\newpage
\thispagestyle{empty}

\begin{center}
{\large Siberian Branch of Russian Academy of Science\\[2mm]
{\normalsize  BUDKER INSTITUTE OF NUCLEAR PHYSICS}
\\[15mm]

%% LaTeX2e file `authors1.tex'
%% generated by the `filecontents' environment
%% from source `rho9495' on 1999/04/23.
%%
R.R.Akhmetshin,  E.V.Anashkin, M.Arpagaus,  V.M.Aulchenko,
V.Sh.Banzarov,  L.M.Barkov,  S.E.Baru,  N.S.Bashtovoy,  A.E.Bondar,
D.V.Chernyak,  A.G.Chertovskikh,  A.S.Dvoretsky, S.I.Eidelman,
G.V.Fedotovich,  N.I.Gabyshev,  A.A.Grebeniuk, D.N.Grigoriev,
B.I.Khazin, I.A.Koop, P.P.Krokovny, L.M.Kurdadze, A.S.Kuzmin,
P.A.Lukin, I.B.Logashenko, A.P.Lysenko, I.N.Nesterenko,
V.S.Okhapkin, E.A.Perevedentsev, A.A.Polunin, E.G.Pozdeev,
V.I.Ptitzyn, T.A.Purlatz, N.I.Root,  A.A.Ruban, N.M.Ryskulov,
A.G.Shamov,  Yu.M.Shatunov,  A.I.Shekhtman, B.A.Shwartz,
V.A.Sidorov,  A.N.Skrinsky,  V.P.Smakhtin, I.G.Snopkov,
E.P.Solodov, P.Yu.Stepanov,  A.I.Sukhanov, V.M.Titov,
Yu.Y.Yudin,  S.G.Zverev,
D.H.Brown,  B.L.Roberts,
J.A.Thompson,
V.W.Hughes

\vskip 15mm
{\Large Measurement of $e^+e^-\rightarrow \pi^+\pi^-$ cross
section with CMD-2 around $\rho$-meson}
\\[15mm]
Budker INP 99-10\\
\vfill
Novosibirsk\\[1mm]
1999 }
\end{center}

\newpage
\thispagestyle{empty}

\begin{center}
\
{\bf Measurement of $e^+e^-\rightarrow \pi^+\pi^-$ cross
section with CMD-2 around $\rho$-meson}
\\[10mm]
%% LaTeX2e file `authors.tex'
%% generated by the `filecontents' environment
%% from source `rho9495' on 1999/04/23.
%%
R.R.Akhmetshin,  E.V.Anashkin, M.Arpagaus,  V.M.Aulchenko,
V.Sh.Banzarov,  L.M.Barkov,  S.E.Baru,  N.S.Bashtovoy,  A.E.Bondar,
D.V.Chernyak,  A.G.Chertovskikh,  A.S.Dvoretsky, S.I.Eidelman,
G.V.Fedotovich,  N.I.Gabyshev,  A.A.Grebeniuk, D.N.Grigoriev,
B.I.Khazin, I.A.Koop, P.P.Krokovny, L.M.Kurdadze, A.S.Kuzmin,
P.A.Lukin, I.B.Logashenko, A.P.Lysenko, I.N.Nesterenko,
V.S.Okhapkin, E.A.Perevedentsev, A.A.Polunin, E.G.Pozdeev,
V.I.Ptitzyn, T.A.Purlatz, N.I.Root,  A.A.Ruban, N.M.Ryskulov,
A.G.Shamov,  Yu.M.Shatunov,  A.I.Shekhtman, B.A.Shwartz,
V.A.Sidorov,  A.N.Skrinsky,  V.P.Smakhtin, I.G.Snopkov,
E.P.Solodov, P.Yu.Stepanov,  A.I.Sukhanov, V.M.Titov,
Yu.Y.Yudin,  S.G.Zverev \\ \vspace{2mm}
{\it Budker Institute of Nuclear Physics, Novosibirsk, 630090,
Russia} \\ \vspace{2mm}
D.H.Brown,  B.L.Roberts \\ \vspace{2mm}
{\it Boston University, Boston, MA 02215, USA} \\ \vspace{2mm}
J.A.Thompson \\ \vspace{2mm}
{\it University of Pittsburgh, Pittsburgh, PA 15260, USA} \\ \vspace{2mm}
V.W.Hughes \\ \vspace{2mm}
{\it Yale University, New Haven, CT 06511, USA}

\end{center}
%% LaTeX2e file `abstract.tex'
%% generated by the `filecontents' environment
%% from source `rho9495' on 1999/04/23.
%%
\begin{center}
{\bf Abstract}\\
\end{center}

{\small
In experiments with the CMD-2 detector  at the VEPP-2M
electron-positron collider at Novosibirsk about 150000 \(
e^{+}e^{-}\rightarrow \pi ^{+}\pi ^{-} \) events  were recorded
in the center-of-mass energy range from 0.61 up to 0.96 GeV. The result
of the pion form factor measurement with a 1.4\% systematic error is
presented. The following values of the $\rho$-meson and
$\rho-\omega$ interference parameters were found: $\mrho=(775.28\pm
0.61\pm 0.20)$ MeV, $\Gamma_\rho=(147.70\pm 1.29 \pm 0.40)$ MeV,
$\Gamma(\rho\rightarrow e^+e^-)=(6.93\pm 0.11\pm 0.10)$ keV,
$Br(\omega\rightarrow\pi^+\pi^-) = ( 1.32\pm 0.23 ) \%$.
}

\vfill
{\normalsize
\copyright \mbox{ } 
\textit{Budker Institute of Nuclear Physics}
}
\newpage

\tableofcontents

\newpage 

\thispagestyle{empty}

\mbox{ }

\newpage 

\section{Introduction}

The cross-section of the process
 $e^{+}e^{-}\rightarrow \pi ^{+}\pi ^{-}$ is given
by
\[
\sigma = \frac{\pi\alpha^2}{3s}\beta^3_\pi \left| F_\pi(s) \right|^2,
\]
where \( F_\pi(s) \) is the pion form factor at the center-of-mass
 energy squared $s$ and $\beta_\pi$ is the pion velocity.

The pion form factor measurement is important for a number of physics
problems. Detailed experimental data in the time-like region allows
measurement of the parameters of the $\rho(770)$ meson and its radial
excitations. Extrapolation of the energy dependence of the pion form
factor to the 
point $s=0$ gives the value of the pion electromagnetic radius. Exact
data on the pion form factor is necessary for precise determination of
the ratio 
\[
R=\sigma (e^{+}e^{-}\rightarrow hadrons)/
\sigma (e^{+}e^{-}\rightarrow \mu ^{+}\mu ^{-}).
\]
 Knowledge of R with high accuracy is required to evaluate the hadronic
contribution $a_{\mu}^{had}$ to the anomalous magnetic moment of the
muon $(g-2)_{\mu}$ \cite{kino}. About 87\% of the hadronic contribution 
in this case comes from $s<2\;\mathrm{GeV^2/c^2}$ (VEPP-2M range), and about
72\% --- from \( e^{+}e^{-}\rightarrow \pi ^{+}\pi ^{-} \) 
channel with \( s<2\; \mathrm{GeV^{2}/c^{2}} \) \cite{ej,bw}. The E821 
experiment at BNL \cite{E821} has
collected its first data in 1997 and will ultimately measure 
$(g-2)_{\mu}$ with a
0.35 ppm accuracy. To calculate the hadronic contribution with the
desired precision 
the systematic error in R should be below 0.5\%. Therefore
a new measurement of the pion form factor with a low systematic error is
required. 

Experiments at the VEPP-2M collider\cite{VEPP}, which started in the early
70s, yielded 
a number of important results in $e^{+}e^{-}$ physics at low 
center-of-mass energies from 360 to 1400 MeV.
The high precision measurement of the pion form factor at VEPP-2M was
done in the late 70s -- early 80s by
OLYA and CMD groups \cite{OLYACMD}. In the CMD experiment, 24 points from 360
to 820 MeV were studied with a systematic uncertainty of about 2\%. In
the OLYA experiment, the energy range from 640 to 1400 MeV was scanned
with small energy steps and the systematic uncertainty varied from 4\%
at the \( \rho  \)-meson peak to 15\% at 1400 MeV. 

\begin{table}
\begin{center}
\begin{tabular}{|c|r|c|c|r|}
\hline 
ID & \multicolumn{1}{|c|}{Date} & $2E$, GeV&
\parbox[t]{2.1cm}{Number of energy points} &
$N_{e^{+}e^{-}\rightarrow \pi ^{+}\pi ^{-}}$\\
\hline 
\hline 
1 & Jan--Feb 1994 & 0.81--1.02 & 14 &   35000 \\ \hline 
2 & Nov--Dec 1994 & 0.78--0.81 & 10 &   66000 \\ \hline 
3 & Mar--Jun 1995 & 0.61--0.79 & 20 &   85000 \\ \hline 
4 & Oct--Nov 1996 & 0.37--0.52 & 10 &    4500 \\ \hline 
5 & Feb--Jun 1997 & 0.98--1.38 & 37 &   75000 \\ \hline 
6 & Mar--Jun 1998 & 0.36--0.97 & 37 & 1900000 \\ \hline 
\end{tabular}
\end{center}
\caption{\label{runstable}CMD-2 runs dedicated to R measurement}
\end{table} 

During 1988-92 a new booster was installed to allow higher positron currents
and injection of the electron and positron beams directly at the
desired energy. 
During 1991-92 a new detector CMD-2 was installed at VEPP-2M, and in 1992 it
started data taking. The pion form factor measurement was one of the
major experiments planned at CMD-2. The energy scan of the whole
VEPP-2M 0.36-1.38 GeV energy range was performed in six separate runs 
listed in Table \ref{runstable}. The energy range below
the $\phi$ meson was scanned twice (in 94-96 and in 98). 

In this article we present results of the analysis of the data from
runs 1--3. The data were taken at 43 energy points with the center-of-mass
energy from 0.61 GeV up to 0.96 GeV with a 0.01 GeV energy step.
The small energy step allows calculation of hadronic contributions in
model-independent way. In the narrow energy region near the
$\omega$-meson the energy steps were $0.002\div 0.006$ GeV in order to
study the $\omega$-meson parameters and the $\rho-\omega$ interference.
 Since the form factor is changing relatively
fast in this energy region, it was important that the beam energy was
measured with the help of the
resonance depolarization technique at almost all energy points. That
allowed a significant decrease of the systematic error coming from
the energy uncertainty.

\begin{figure}
\begin{center}
\begin{tabular}{cc}
\includegraphics[width=0.45\textwidth]{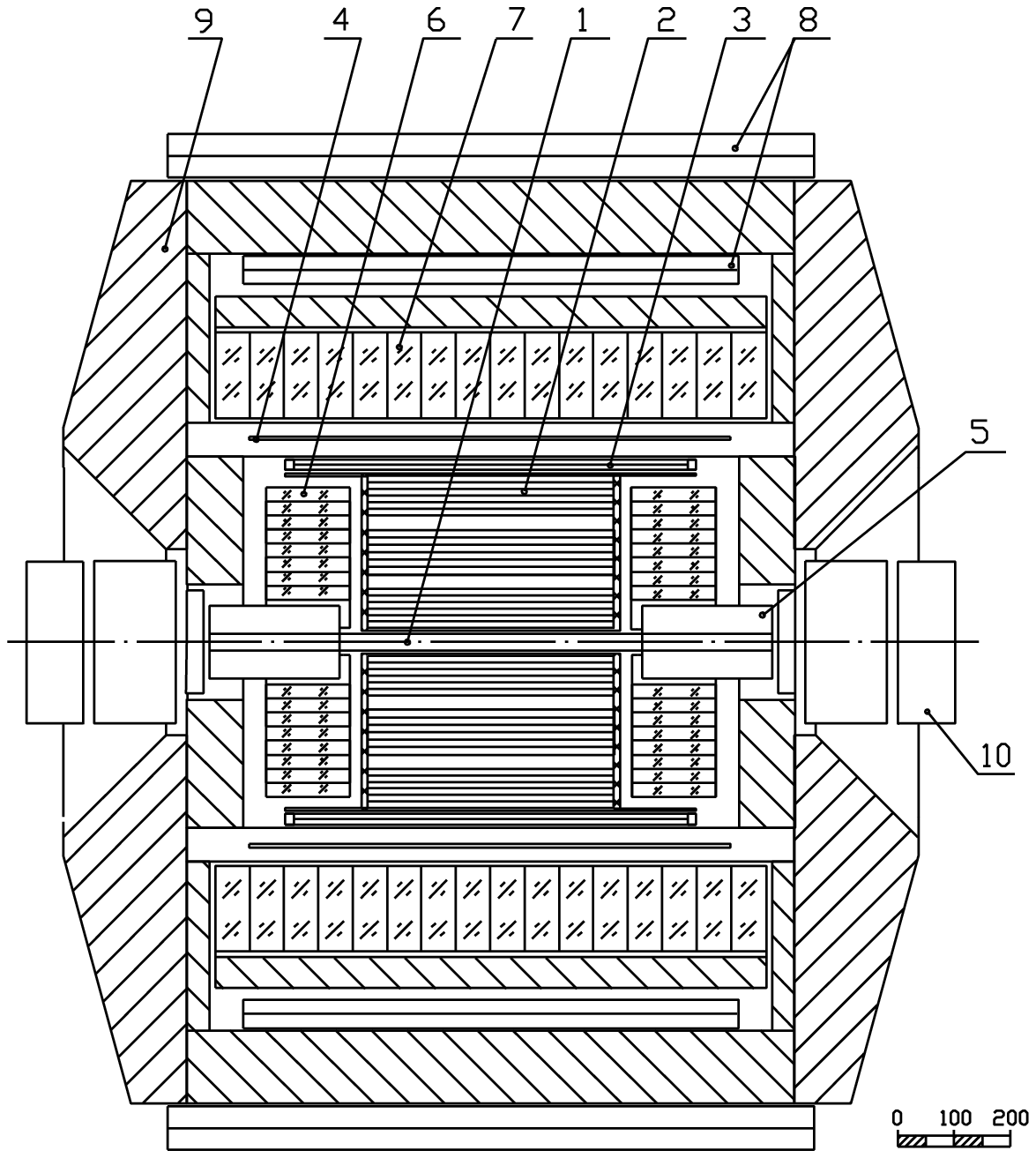}  & 
\includegraphics[width=0.5\textwidth]{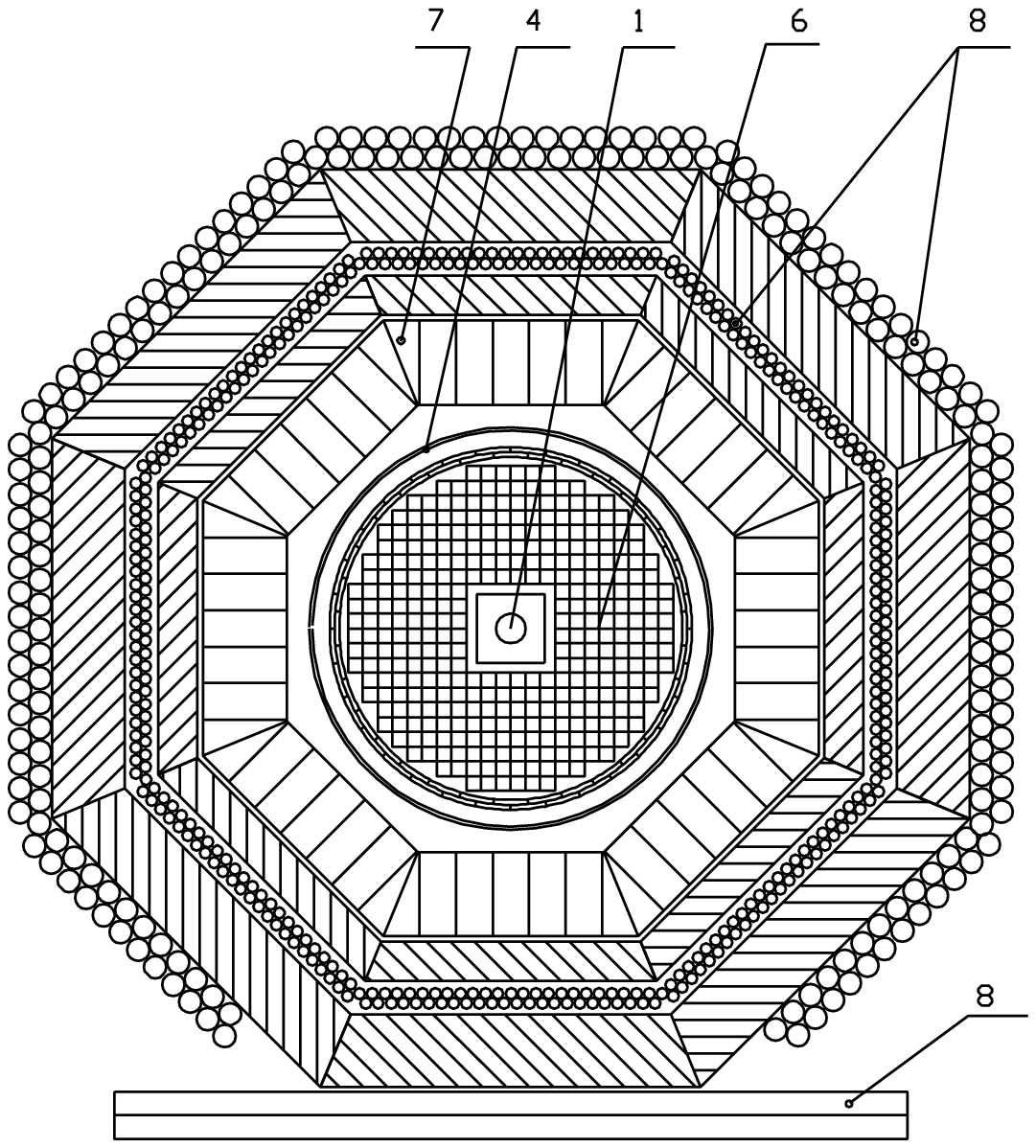}\\
\end{tabular}
\end{center}
\caption{\label{cmd2p}Detector CMD-2. 1 --- beam pipe, 2 --- drift
chamber, 3 --- Z-chamber, 4 --- main solenoid, 5 --- compensating
solenoid, 6 --- endcap (BGO) calorimeter, 7 --- barrel (CsI)
calorimeter, 8 --- muon range system, 9 --- magnet yoke, 10 --- 
storage ring lenses}
\end{figure}

The CMD-2 (Fig.\ \ref{cmd2p}) is a general purpose detector consisting
of the drift chamber, the proportional Z-chamber, the barrel (CsI) and the
endcap (BGO) electromagnetic calorimeters and the muon range system. The
drift chamber, Z-chamber and the endcap calorimeters are installed 
inside a thin superconducting solenoid with a field of 10 kGs. 
More details on the detector can be found elsewhere
\cite{ICFA,CMD2,PREP}.
The data described here was taken before the endcap calorimeter was
installed. 

Two independent triggers were used during data taking. The first one, {\em
charged trigger}, analyses information from the drift chamber and the
Z-chamber and triggers the  detector if at least one track was
found. For 0.81--0.96 GeV energy points there was an additional
requirement for the
total energy deposition in the calorimeter to be 
greater than a 20--30 MeV threshold.  The second one, {\em neutral trigger},
triggers the detector according to information from the calorimeter
only. Events triggered by the charged trigger were used for
analysis, while events triggered by the neutral trigger were used for
trigger efficiency monitoring. 

%\newpage

\section{Data analysis}

\subsection{Selection of collinear events}

\begin{figure}
\begin{center}
\includegraphics[width=0.9\textwidth]{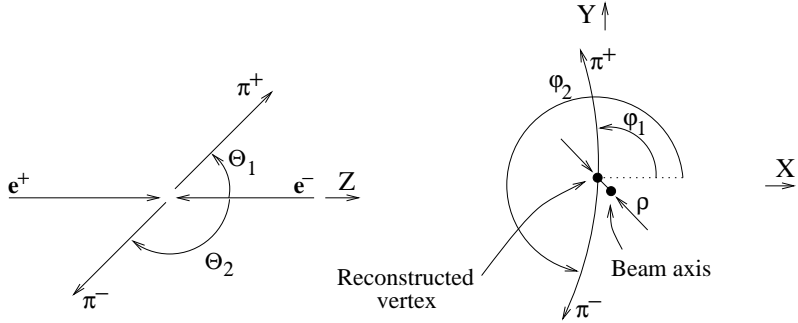}
\end{center}
\caption{\label{colle}Definition of parameters for two track events}
\end{figure}

The data was collected at 43 energy points. At the beam energy of 405
 MeV the data was collected in two different runs with
different detector and trigger conditions. Therefore, two
independent results for this energy point are presented.

From more than \( 4\cdot 10^{7} \) triggers about \( 4\cdot 10^{5} \) 
were selected as collinear events. The selection criteria were as
follows.\label{collsel} 

\begin{enumerate}
\item The event was triggered by the charged trigger. Other triggers may
be present as well.
\item Only one vertex with two oppositely charged
tracks was found in the drift chamber.
\item Distance from the vertex to the beam axis $\rho$ is less than 0.3 cm.
\item Z-coordinate of the vertex (distance to the interaction point
along the beam axis) $|Z|$ is less than 8 cm.
\item Average momentum of two tracks $(p_1+p_2)/2$ is between 200 and
600 MeV/c.
\item Acollinearity of two tracks in the plane transverse to the beam
axis $ |\Delta \varphi |=|\pi -|\varphi _{1}-\varphi _{2}|| $
is less than 0.15 radians. 
\item Acollinearity of two tracks in the plane that contains the beam
axis $|\Delta \Theta |=|\Theta _{1}-(\pi -\Theta _{2})|$ 
is less than 0.25 radians.
\item Average polar angle of two tracks $ [\Theta _{1}+(\pi -\Theta _{2})]/2 $
is between $\Theta _{min}$ and $(\pi -\Theta _{min})$. All analysis
was done separately for $\Theta _{min}=1.0$ and $\Theta _{min}=1.1$ radian.
\end{enumerate}

Because of the unfortunate accident the high voltage was off for two
neighbouring lines of the
CsI calorimeter during the data taking in the 0.61--0.784 GeV energy
region. In order to avoid additional calorimeter inefficiency, 
events with $4.35<\varphi_{+}<4.95$ 
or $3.90<\varphi _{-}<4.50$ have been removed from analysis for
corresponding energy points (angles are measured in radians). Events
with $1.05<\varphi _{+}<1.65$ or 
$0.55<\varphi _{-}<1.15$ were additionally removed for 0.65--0.70 GeV
energy points. 

%\begin{figure}
%\begin{center}
%\begin{tabular}{cc}
%\subfigure[$\Delta\varphi=\pi-|\varphi_1-\varphi_2|$.]
%{\includegraphics[width=0.45\textwidth]{rho9495_sel_1.eps}}  & 
%\subfigure[$\Delta\Theta=\pi-(\Theta_1+\Theta_2)$.]
%{\includegraphics[width=0.45\textwidth]{rho9495_sel_2.eps}}  \\
%\subfigure[$\rho$.]
%{\includegraphics[width=0.45\textwidth]{rho9495_sel_3.eps}}  & 
%\subfigure[$Z$.]
%{\includegraphics[width=0.45\textwidth]{rho9495_sel_4.eps}}  \\
%\subfigure[$P_{AVR}=(P_1+P_2)/2$.]
%{\includegraphics[width=0.45\textwidth]{rho9495_sel_5.eps}}  & 
%\subfigure[$\Theta_{AVR}=(\Theta_1+\pi-\Theta_2)/2$.]
%{\includegraphics[width=0.45\textwidth]{rho9495_sel_6.eps}}  
%\end{tabular}
%\end{center}
%\caption{\label{selp} Selection of the collinear events. Empty
%histograms correspond to real collinear events, dashed histogram
%correspond to simulated $e^+e^-\rightarrow e^+e^-(\gamma)$
%events. Both data and simulation correspond to 0.782 GeV energy point.
%The simulation and the real data histograms have the different number
%of events.
%}
%\end{figure}

Definition of the collinear event parameters, such as \( \rho  \), \(
\Theta _{1,2} \), 
\( \varphi _{1,2} \), is illustrated in Fig.\ \ref{colle}.

\subsection{Event separation}

\subsubsection{Likelihood function}

\begin{floatingfigure}{0.46\textwidth}
\begin{center}
\includegraphics*[width=0.45\textwidth]{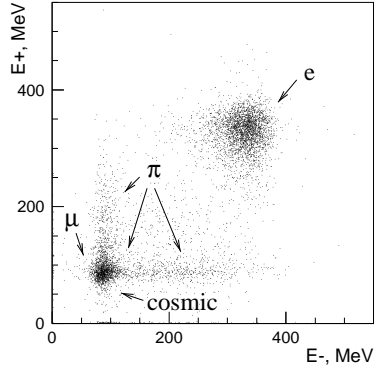} 
\end{center}
\caption{\label{distr}$E^+$ versus $E^-$ distribution
for collinear events for 0.8 GeV energy point\vspace{2mm}}
\end{floatingfigure}

The selected sample of events consists of collinear events 
\( e^{+}e^{_{-}}\rightarrow e^{+}e^{-}, \) 
\( e^{+}e^{_{-}}\rightarrow \pi ^{+}\pi ^{-} \), 
\( e^{+}e^{_{-}}\rightarrow \mu ^{+}\mu ^{_{-}} \)
and background events mainly due to cosmic muons. 
The energy deposition in the barrel CsI calorimeter
by negati\-vely $(E^-)$
and positively $(E^+)$ charged particles  were
used for event separation.
The distribution of \( E^{+} \) versus \( E^{-} \) is shown in
Fig.\ \ref{distr}. Electrons and positrons usually have  large energy
deposition since they produce electromagnetic showers. Muons usually 
have small
energy deposition since they are minimum ionising particles. Pions can
interact as minimum ionising particles producing small energy
deposition or  have nuclear interactions inside the calo\-ri\-meter,
resulting in long tails to a higher value of the energy deposition.

The separation was based on the minimization of the following unbinned likelihood function:
\begin{equation}
\label{lfunc1}
L=-\sum _{events}\ln \left( \sum _{a}N_{a}\cdot f_{a}(E^{+},E^{-})\right) +\sum _{a}N_{a},
\end{equation}
where \( a \) is the event type ($a=ee$, $\mu\mu$, $\pi\pi$, $cosmic$),
\( N_{a} \) is the number of events of the type \( a \) and \( f_{a}(E^{+},E^{-}) \)
is the probability density for a type \( a \) event to have energy depositions
\( E^{+} \) and \( E^{-} \).
It was assumed that \( E^{+} \) is uncorrelated with  \( E^{-} \)
for events of the same type, so we can factorize the probability density:
\[
f_{a}(E^{+},E^{-})=f_{a}^{+}(E^{+})\cdot f^{-}_{a}(E^{-}).\]
 For \( e^{+}e^{-} \), \( \mu ^{+}\mu ^{-} \) pairs and cosmic events the
energy deposition distribution is the same for negatively and positively charged
particles, while these distributions are significantly different for \( \pi ^{+} \)
and \( \pi ^{-} \). Therefore \( f_{a}^{+}\equiv f^{-}_{a} \) for \( a=ee,\; \mu \mu  \)
and \( cosmic \), but for pions these functions are different. The probability
density functions are described in detail in the following sections. 

In minimization the ratio \( N_{\mu \mu }/N_{ee} \) was fixed according
to the QED calculation
\[
\frac{N_{\mu\mu}}{N_{ee}}=
\frac{\sigma_{\mu\mu}\cdot (1+\delta_{\mu\mu})
\left( 1+\alpha_{\mu\mu} \right) \varepsilon_{\mu\mu}}
{\sigma_{ee}\cdot (1+\delta_{ee}) \left( 1+\alpha_{ee} \right) 
\varepsilon_{ee}},
\]
 where $\sigma$ is the Born cross-section, $\delta$ is a
 radiative correction, 
\( \alpha  \) is the correction for the experimental resolution of $\Theta$
angle measurement and \( \varepsilon  \) is the reconstruction efficiency (see
sec. \ref{piformsec} for details). The likelihood function (\ref{lfunc1})
was rewritten to have the following global fit parameters:
\[
(N_{ee}+N_{\mu \mu }),\quad \frac{N_{\pi \pi }}{N_{ee}+N_{\mu \mu }},\quad N_{cosmic}\]
 instead of \( N_{ee} \), \( N_{\mu \mu } \), \( N_{\pi \pi } \) and \( N_{cosmic} \).
The number of cosmic events $N_{cosmic}$ was determined before the fit as
described below, and was fixed during the fit.
\vspace{2mm}

\subsubsection{Rejection of background events}

\begin{floatingfigure}{0.47\textwidth}
\begin{center}
\includegraphics*[width=0.4\textwidth]{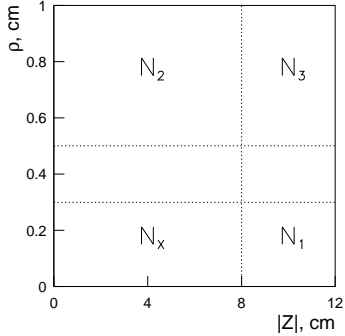} 
\end{center}
\caption{\label{rhoz} To determination of number of cosmic background
events}
\end{floatingfigure}

\begin{figure}[tb]
\begin{center}
\begin{tabular}{cc}
\subfigure[\label{rhoz1}$\rho$-distribution]
{\includegraphics[width=0.45\textwidth]{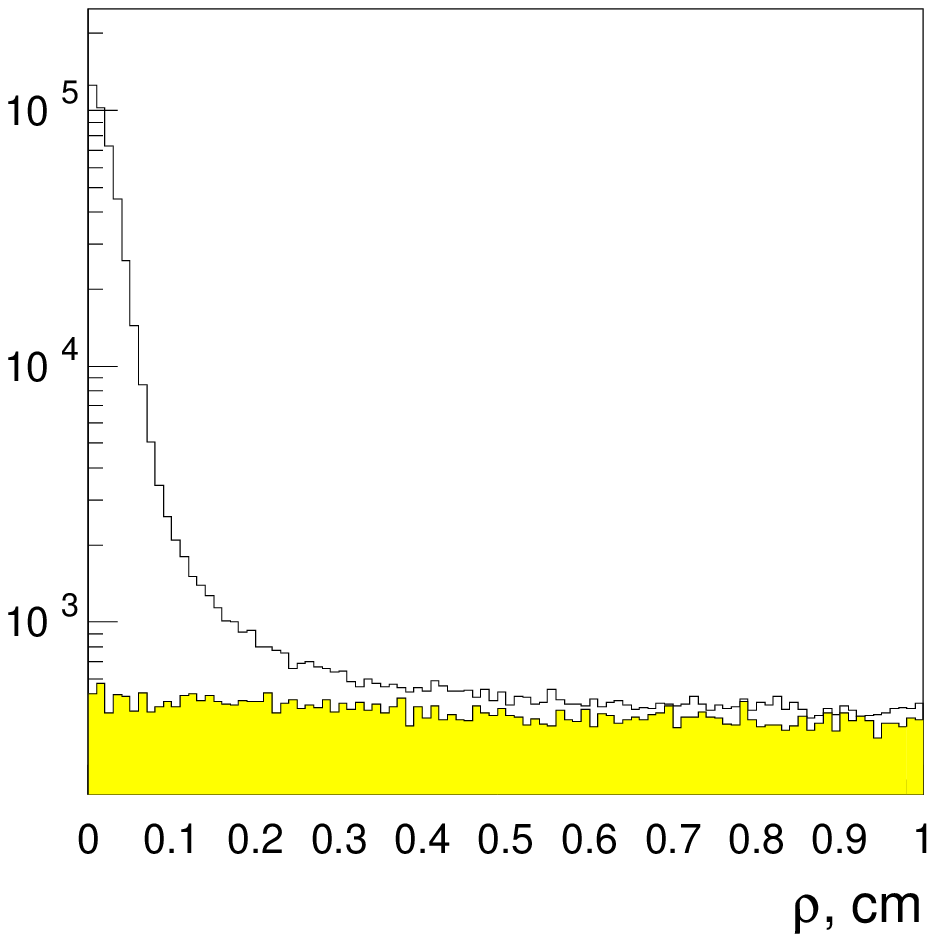}} & 
\subfigure[\label{rhoz2}$Z$-distribution]
{\includegraphics[width=0.45\textwidth]{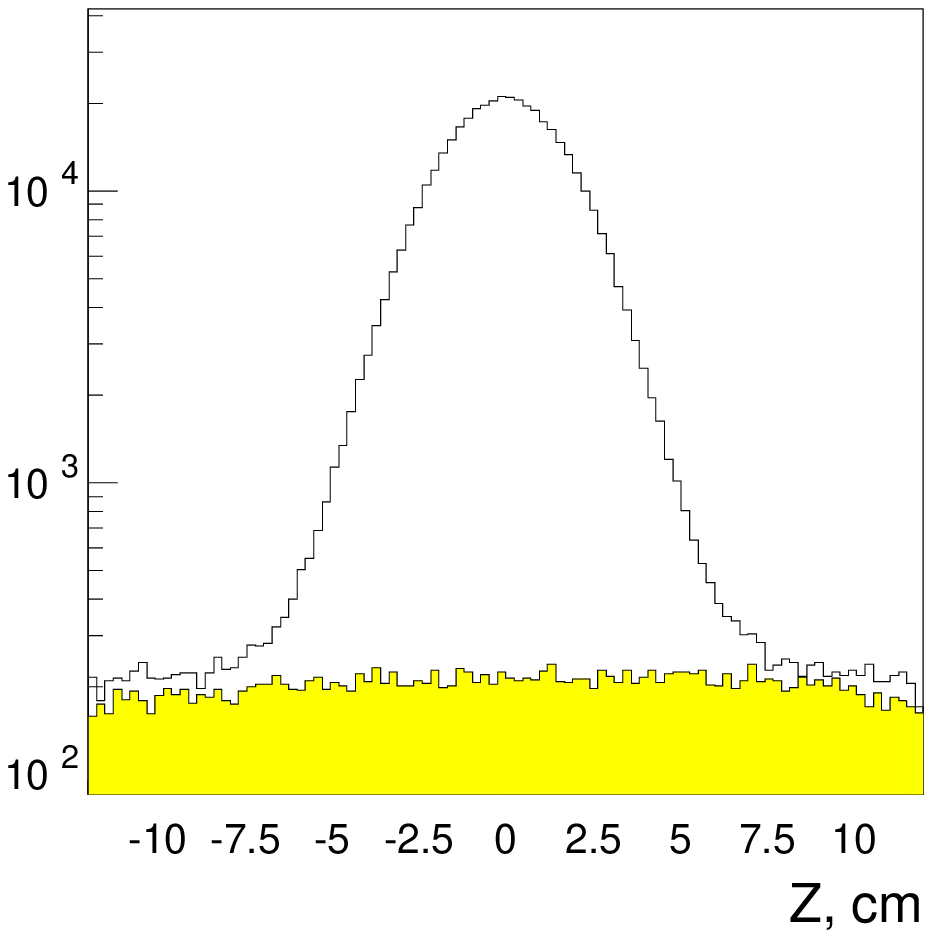}} \\
\end{tabular}
\end{center}
\caption{\label{rhoz12} $\rho$- and $Z$-distributions for background
and collinear events. Empty histograms correspond to all selected
events, filled histograms correspond to background events only}
\end{figure}

Cosmic background events can be separated from the beam-produced col\-linear
events by the distance $\rho$ from the vertex to beam axis and (or) by
the $Z$-coordinate of the vertex. The $\rho$-distribution for collinear
events has a narrow peak around 0, while the same distribution for the
background events is almost flat (Fig.\ \ref{rhoz1}). Similarly, the
$Z$-distribution for 
collinear events is Gaussian like, while it is almost flat for the
background events (Fig.\ \ref{rhoz2}). 
Assuming that $\rho$- and $Z$-distributions for cosmic events are not
correlated and that all events with $\rho>0.5$ cm or $|Z|>8$ cm are
cosmic events, one can calculate the number of background events in the
signal region from the  number of events out of the signal region. 

Let us define $N_X$ as the number of cosmic events in the signal region
($\rho<0.3$ cm, $|Z|\leq 8$ cm) and $N_{1,2,3}$ as the number of
events in different out-of-signal $\rho-Z$ regions (see Fig.\ \ref{rhoz}):
\[
\left\{
\begin{array}{l}
N_1=N\bigm|_{\rho<0.3;\;8<|Z|<12} \\
N_2=N\bigm|_{0.5<\rho<1.0;\;|Z|\leq 8} \\
N_3=N\bigm|_{0.5<\rho<1.0;\;8<|Z|<12}
\end{array}
\right.
.
\]
Then under assumptions mentioned above we can derive:
\[ N_X = \frac{N_2}{N_3}\cdot N_1 = \frac{N_1}{N_3}\cdot N_2. \]
$N_X$ is equal to $N_{cosmic}$ only statistically. The
confidence interval for $N_{cosmic}$, corresponding to one standard
deviation, is
\[
N_{cosmic}
%= N_X \pm N_X \cdot
%\sqrt{\frac{1}{N_1}+\frac{1}{N_2}+\frac{1}{N_3}+\frac{1}{N_X}}
\approx N_X \pm \sqrt{5 N_X}.
\]
In order to take that into account, the
$N_{cosmic}$ value was fixed to the $N_X$ value during fit, but 
$\Delta N_X$ was added to $\Delta N_{\pi\pi}$ at the end.

\vspace{2mm}

\subsubsection{Energy deposition parametrization}

In order to construct the likelihood function (\ref{lfunc1}) it is very
convenient to have an analytical form for energy deposition
distributions. We parametrize all energy depositions by the linear
combination of two types of Gaussian-like functions, described below. 

\begin{description}
\item [Normal distribution]\( g(x;x_{0},\sigma ) \) 
\[
g(x;x_{0},\sigma )=\frac{1}{\sqrt{2\pi }\sigma }\exp \left[ -(x-x_{0})^{2}/2\sigma ^{2}\right] ,\]
 where the parameters are the mean value $x_0$ and 
the standard deviation $\sigma$. The Gaussian is normalized at the
$(-\infty, +\infty)$ interval, 
but energy deposition values belong to the $[0,+\infty)$ range. Therefore
for parametrization the Gaussian normalized at the $[0,+\infty)$
range is used:
\[
g(x;x_{0},\sigma )/\left[ 1-I\left( -\frac{x_{0}}{\sigma }\right)
\right] ,
\]
 where 
\[
I(x)=\frac{1}{\sqrt{2\pi }}\int _{-\infty }^{x}\exp \left(
-\frac{t^{2}}{2}\right) dt.
\]
 
\item [Normal logarithmic distribution\cite{lgs}]
 \( g_{l}(x;x_{0},\sigma ,\eta ) \)
\[
g_{l}(x;x_{0},\sigma ,\eta )=\frac{1}{\sqrt{2\pi }\sigma }\cdot
 \frac{\eta }{\sigma _{0}}\cdot \exp \left\{ -\frac{1}{2}\left[
 \frac{\ln ^{2}\left( 1-\frac{x-x_{0}}{\sigma }\eta \right) }{\sigma
 _{0}^{2}}+\sigma _{0}^{2}\right] \right\} ,\] 
 where 
\[ \sigma_0=\ln\left(\eta\sqrt{2\ln 2}+\sqrt{1+\eta^2\cdot 2\ln 2}
   \right) \left/ \sqrt{2\ln 2} \right. . \]
 Parameters of this function are the most probable energy $x_0$, the %\linebreak
 $\sigma=\mathrm{FWHM}/2.35$ and the asymmetry $\eta$. If \( \eta =0 \), the logarithmic Gaussian is equal to the plain Gaussian:
\[
g_{l}(x;x_{0,}\sigma ,\eta )\left| _{\eta =0}\equiv g(x;x_{0},\sigma
).\right. 
\]
The function is normalized
 at the \( [-\infty ,x_{0}+\sigma /\eta ] \) interval:
\[
\int _{-\infty }^{x_{0}+\sigma /\eta }g_{l}(x)dx=1,\]
 but in our case its integral over the range \( (-\infty ,0] \) is negligible
and it is assumed that this function is normalized at the \( [0,x_{0}+\sigma /\eta ] \)
interval. 
\end{description}

Since both functions are normalized,  it is easy to construct the
normalized energy deposition 
distribution. 

\vspace{2mm}

\subsubsection{\label{ecorsec}Correction of energy deposition for polar angle}

\begin{figure}
\begin{center}
\begin{tabular}{cc}
\subfigure[Energy deposition versus $\Theta$ for collinear events before correction]{\resizebox*{0.4\textwidth}{!}{\includegraphics{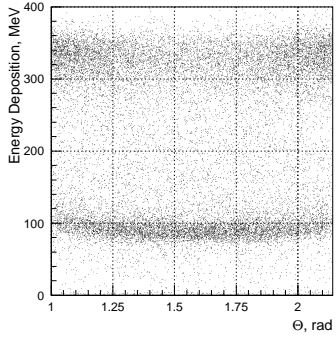}}} &
\subfigure[Average energy deposition versus $\Theta$ for electrons and
cosmic background events
 before correction]{\resizebox*{0.4\textwidth}{!}{\includegraphics{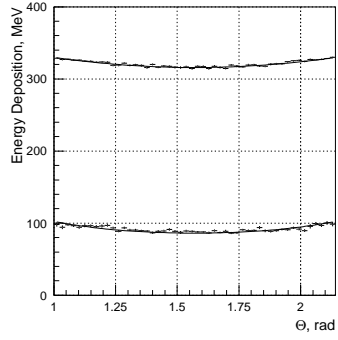}}} \\
\subfigure[Energy deposition versus $\Theta$ for collinear events after correction]{\resizebox*{0.4\textwidth}{!}{\includegraphics{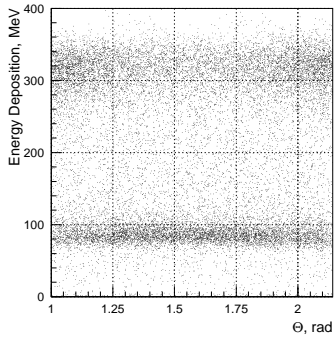}}} &
\subfigure[Average energy deposition versus $\Theta$ for electrons and
cosmic background events after correction]
{\resizebox*{0.4\textwidth}{!}{\includegraphics{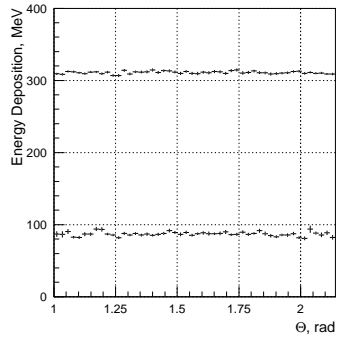}}}
\\ 
\end{tabular}
\end{center}
\caption{\label{ecorfig}Correction of energy deposition for the
 polar angle of the track}
\end{figure}

Energy deposition in the calorimeter is correlated with the polar angle
$\Theta$
of a track due to the dependence of calorimeter thickness on $\Theta$. So the
probability density for an event of type $a$ to have energy deposition $E$
depends not only on $E$, but also on $\Theta$. In order to simplify
the likelihood function, the energy deposition was corrected to make
dependence of energy deposition on \( \Theta  \) negligible. In other words,
instead of \( f_{a}(E,\Theta ) \) we are using \( f_{a}(\overline{E}) \),
where \( \overline{E} \) is the corrected energy deposition:
\[
E\longrightarrow \overline{E}\quad to\; have\quad f_{a}(E,\Theta )\longrightarrow f_{a}(\overline{E}).\]

The correction procedure was performed as follows. First, two groups of events were
selected: cosmic events and electrons. Selection criteria for both types of
events are described below in sections \ref{cosesec} and
\ref{elecesec}. 
The scatter-plot
of the energy deposition versus \( \Theta  \) for collinear events is shown
in Fig.\ \ref{ecorfig}a. The full range  of possible \( \Theta  \)
values \( \left[ 1,\pi -1\right]  \) 
was divided into 50 equal bins and for each bin an average energy
deposition  was calculated for electrons and cosmic events. 

The energy deposition of electrons depends on $\Theta$ because
the CMD-2 calorimeter has only about $8X_{0}$ and therefore the shower
leakage is not negligible. The effective calorimeter thickness
increases for incident track angle away from normal.
The corresponding small variation of the average energy deposition of
electrons was fit to the parabolic function
\[E_e(\Theta)=E_1\cdot \left[ 1+\alpha (\Theta-\pi/2)^2\right],\]
where $E_{1}$ and $\alpha$ are the fit parameters. 

Cosmic events contain minimum ionising particles and their energy deposition is
proportional to the length in the calorimeter through which the
particle passes. This length is proportional to \( 1/\sin \Theta  \) ,
and the fit function in this case was 
\[E_{c}(\Theta )=E_{0}/\sin \Theta ,\]
where $E_{0}$ is a fit parameter. An example of the fit of the average
energy depositions for electrons and cosmic tracks is shown in 
Fig.~\ref{ecorfig}b.

The corrected energy was calculated as 
\[
\overline{E}(E,\Theta )=E\cdot \left[ k_{c}(\Theta )+\frac{k_{e}(\Theta )-k_{c}(\Theta )}{E_{e}(\Theta )-E_{c}(\Theta )}\left( E-E_{c}(\Theta )\right) \right] ,\]
 where \( k_{c}(\Theta )=E_{0}/E_{c}(\Theta ) \), \( k_{e}(\Theta )=E_{1}/E_{e}(\Theta ) \),
\( E_{c}(\Theta ) \) and \( E_{e}(\Theta ) \) are the measured average
energy deposition for cosmic and Bhabha events respectively. 
After such a correction the average
energy deposition for both cosmic and Bhabha events does not depend on \( \Theta  \).
In fact, if \( E=E_{e}(\Theta ) \), then \( \overline{E}=E_{1} \); if \( E=E_{c}(\Theta ) \),
then \( \overline{E}=E_{0} \).

Since the energy deposition of cosmic events does not depend on the beam
energy, the variations of $E_0$ reflects residual variations in
the calibration of the CsI calorimeter. To correct for these variations,
the energy deposition was normalized  to have the average
energy deposition of the cosmic events equal to 85 MeV (the arbitrary
constant close to the experimental value) for all beam energies:
\[
\overline{E}\longrightarrow \overline{E}\cdot \frac{85}{E_{0}}.\]

The scatter-plot of the corrected energy deposition versus \( \Theta  \) for collinear
events is shown in Fig.\ \ref{ecorfig}c. The average energy deposition for electrons
and cosmic events versus \( \Theta  \) after the correction is shown in Fig.\ \ref{ecorfig}d. 

The described energy deposition correction was calculated independently for each energy
point. In the all following sections \( E \) actually means the
corrected energy  $\overline{E}$.

\vspace{2mm}

\subsubsection{\label{cosesec}Energy deposition of cosmic events}

As mentioned above, it is possible to select a pure sample of
background events from the real data. In this case the selection criteria
are the same as standard cuts for the collinear events, except that
the distance $\rho$ from the vertex to the beam axis should be from 5 to
10 mm (instead of less than 3 mm). In order to make the selected sample of
background events even cleaner, the fact that there are two clusters in
the event was used. Analysing the energy deposition of one cluster,
the energy deposition of the other one was required to be less than 150
MeV. Since the clusters are independent,
the strict limit on one cluster energy does not affect the energy
deposition distribution for the other cluster.

\begin{figure}[tbp]
\begin{center}
\includegraphics[width=0.9\textwidth]{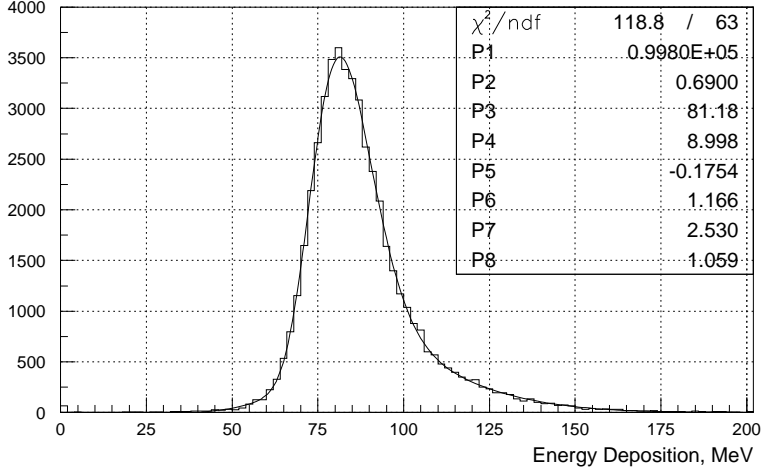}
\end{center}
\caption{\label{cosfit}Fit of the energy deposition for cosmic events with
the function (\ref{fcosmic1}).
Fit parameters: P1 -- number of events, P2 -- $p$, P3 -- $E_0$,
P4 -- $\sigma$, P5 -- $\eta$, P6 -- $\alpha$, P7 -- $\beta$, 
P8 -- $\gamma$}
\end{figure}

The data for all energy points was combined together and the 
energy deposition of cosmic events was parametrized using the following
function: 
\begin{equation}
\label{fcosmic1}
f(E)=p\cdot g_{l}(E;E_{0},\sigma ,\eta )+(1-p)\cdot g_{l}(E;E_{0}+\alpha \sigma ,\beta \sigma ,\gamma \eta ).
\end{equation}
The fit result is presented in Fig.\ \ref{cosfit}. The obtained
function was used as the energy deposition for background events:
\begin{equation}
\label{fcosmic0}
\begin{array}{c}
f^{+}_{cosmic}(E)\equiv f^{-}_{cosmic}(E)\equiv f_{cosmic}(E) ,
\vphantom{\Big /} \\
\multicolumn{1}{l}{
f_{cosmic}(E) = 0.69\cdot g_{l}(E;81.2,9.0,-0.175)+
\vphantom{\Big /} } \\
\multicolumn{1}{r}{ \qquad \qquad \qquad \qquad \qquad \qquad
+ 0.31\cdot g_{l}(E;91.7,22.8,-0.186).
\vphantom{\Big /} }
\end{array}
\end{equation}

\begin{figure}[p]
\begin{center}
\begin{tabular}{cc}
\subfigure[Group 0.61--0.70 GeV]
{\includegraphics[width=0.45\textwidth]{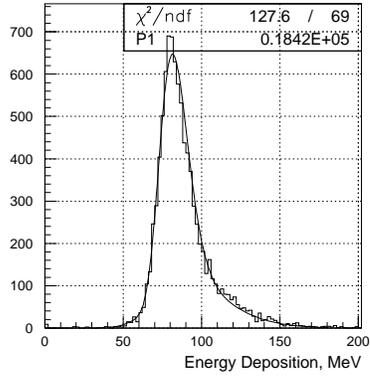}} &
\subfigure[Group 0.71--0.778 GeV]
{\includegraphics[width=0.45\textwidth]{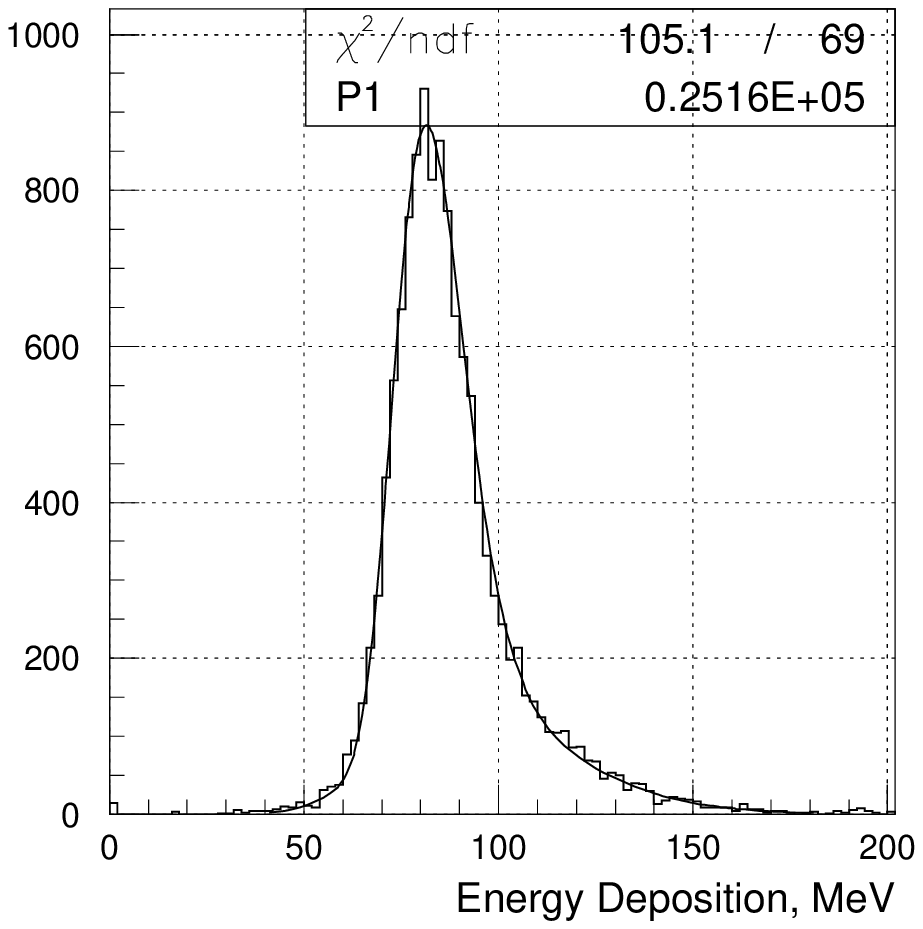}} \\
\subfigure[Group 0.780--0.810 GeV]
{\includegraphics[width=0.45\textwidth]{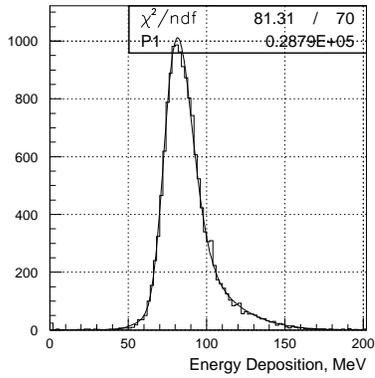}} &
\subfigure[Group 0.810--0.960 GeV]
{\includegraphics[width=0.45\textwidth]{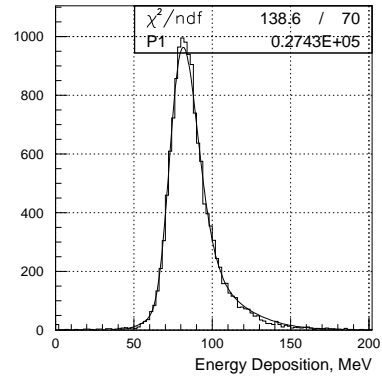}} \\
\end{tabular}
\end{center}
\caption{\label{cosefitg}Fit of energy deposition of cosmic events
with the function (\ref{fcosmic0}).
Data is combined into 4 groups of energy points. Fit parameter: P1 -- number
of events}
\end{figure}

In order to check the stability of the energy deposition for cosmic
events, all energy points were combined into 4 groups: 
0.61--0.70 GeV, 0.71--0.778 GeV, 0.780--0.810 GeV and 0.810--0.960 GeV.
The energy deposition of the selected
background events for each group was fitted with the function 
(\ref{fcosmic0}). Results of the fit are demonstrated
in Fig.\ \ref{cosefitg}.

The probability to have no cluster in the calorimeter (zero energy deposition)
for cosmic event tracks was also measured and found to be negligible --- less
than 0.1\%.

\vspace{2mm}

\subsubsection{\label{elecesec}Energy deposition of electrons and positrons}

\begin{figure}[tb]
{\centering \begin{tabular}{cc}
\subfigure[Fit of the positron energy deposition]{\resizebox*{0.46\textwidth}{!}{\includegraphics{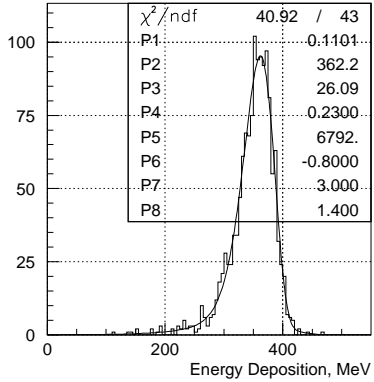}}} &
\subfigure[Fit of the electron energy deposition]{\resizebox*{0.46\textwidth}{!}{\includegraphics{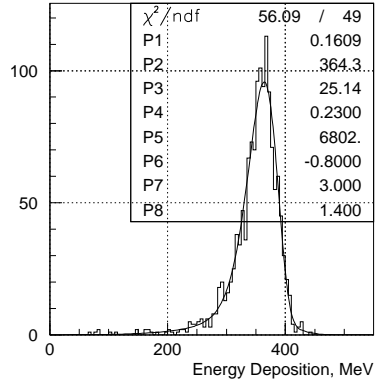}}} \\
\end{tabular}\par}
\caption{\label{elecefit}Fit of the energy deposition of electrons and
positrons at center-of-mass energy 0.9 GeV. 
Fit parameters: P1 -- $p_{w}$, P2 -- $E_{0}$, P3 -- $\sigma$, P4 --
$\eta$, P5 -- number of events, P6 -- $\alpha$, P7 -- $\beta$,
P8 -- $\gamma$}
\end{figure}

The energy deposition for electrons and positrons was obtained from the
data. Electrons and positrons can be separated from mesons by
their relatively high energy deposition in the calorimeter. If we have one
``clean'' electron (positron) in the event (a particle with a proper
momentum and high energy deposition), then we know that the second
particle is a positron (electron), and therefore may be used for the
energy deposition study. 

The test $e^+e^-\rightarrow e^+e^-$ events were selected for the
energy deposition study 
with the same selection criteria as for collinear events except for
the following.
\begin{enumerate}
\item The distance $\rho$ from the vertex to the beam axis is less
than 0.15 cm (instead of 0.3 cm).
\item The average momentum of the particles is within the 10 MeV/c
range around the 
beam energy. This requirement allows to decrease significantly a small
admixture of $e^+e^- \rightarrow \pi^+\pi^-$ events.
\end{enumerate}

Analysing the energy deposition in one cluster, the energy deposition
in the other one is required to be between $0.92E_B-100$ MeV and $0.92E_B$,
where $E_B$ is the beam energy. As for cosmic rays since the clusters
are independent, 
the strict limit on one cluster energy does not affect the energy
deposition distribution for the other cluster.

For each energy point the energy depositions of the selected electrons
and positrons were fitted with the following function:
\[
f(E)=(1-p_{w})\cdot g_{l}(E;E_{0},\sigma ,\eta )+p_{w}\cdot g_{l}(E;E_{0}+\alpha \sigma ,\beta \sigma ,\gamma \eta ).\]
After analysis of the fit results the following values of the
parameters were found:
\[ \alpha =-0.8,\quad \beta =3.0,\quad \gamma =1.4.\]
Again, as for cosmic rays, the probability to have no cluster 
was found to be negligible (less than 0.1\%) for all 
energy points. Finally, the energy deposition of electrons and
positrons was parametrized as:
\begin{equation}
\label{fee0}
\begin{array}{c}
f^{+}_{ee}(E)\equiv f^{-}_{ee}(E)\equiv f_{ee}(E),\\
\\
f_{ee}(E)=(1-p_{w})\cdot g_{l}(E;E_{0},\sigma ,\eta)+p_{w}\cdot g_{l}(E;E_{0}-0.8\sigma ,3\sigma ,1.4\eta),
\end{array}
\end{equation}
 where \( p_{w} \), \( E_{0} \), \( \sigma  \) and $\eta$ are free
parameters of the fit.
%Results of the fit of the energy deposition of electrons and positrons
%for all energy points are shown in Fig.\ \ref{elecefitpar}. 
An example of the electron and
positron energy depositions is shown in Fig.\ \ref{elecefit}.

%\begin{figure}
%{\centering \begin{tabular}{cc}
%\subfigure[$p_W$]{\resizebox*{0.4\textwidth}{!}{\includegraphics{epar1.eps}}} &
%\subfigure[$E_0$]{\resizebox*{0.4\textwidth}{!}{\includegraphics{epar2.eps}}} \\
%\subfigure[$\sigma$]{\resizebox*{0.4\textwidth}{!}{\includegraphics{epar3.eps}}} &
%\subfigure[$\chi^2$ of the fit]{\resizebox*{0.4\textwidth}{!}{\includegraphics{epar4.eps}}} \\
%\end{tabular}\par}
%\caption{\label{elecefitpar}Results of the fit of electron and positron
%energy depositions for all energy points. }
%\end{figure}

\vspace{2mm}

\subsubsection{Energy deposition of muons}

\begin{figure}
\begin{center}
\begin{tabular}{cc}
\multicolumn{2}{c}
{\subfigure[Energy deposition of 375 MeV muons]
{\includegraphics[width=0.95\textwidth]{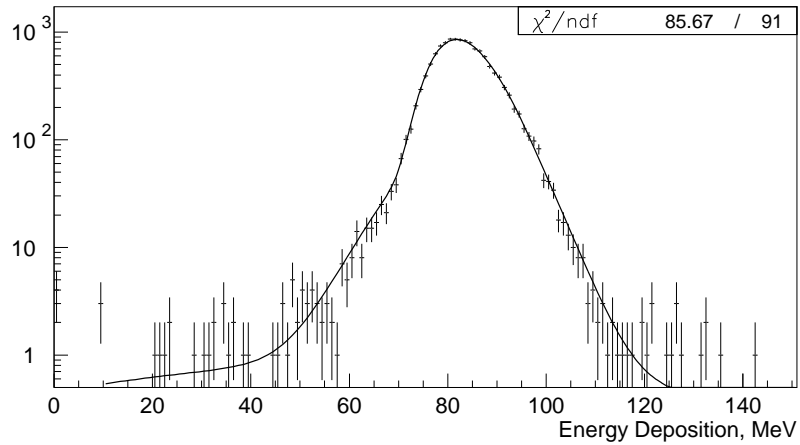}}} \\
\subfigure[Energy dependence of $E_0$]
{\includegraphics[width=0.46\textwidth]{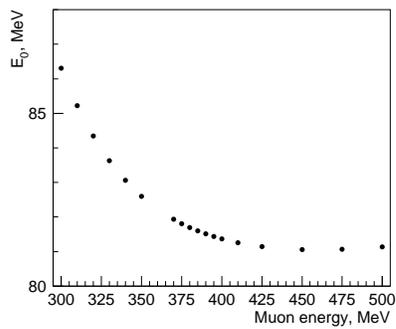}} &
\subfigure[Energy dependence of $\sigma_0$]
{\includegraphics[width=0.46\textwidth]{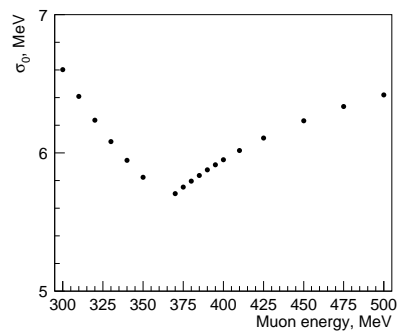}} \\
\end{tabular}
\end{center}
\caption{\label{mip_mu_sim} Simulated energy deposition of
muons}
\end{figure}

In the center-of-mass energy range above 0.6 GeV, muons produced in the reaction 
$e^{+}e^{-}\rightarrow \mu ^{+}\mu ^{-}$
interact with the calorimeter as the minimum ionising
particles,
and therefore their energy deposition can be obtained from the 
detector simulation. The simulation of the $e^+e^-\rightarrow
\mu^+\mu^-$ events was performed for center-of-mass energies 0.6, 0.62,
0.64, 0.66, 0.68, 0.70, 0.74, 0.75, 0.76, 0.77, 0.78, 0.79, 0.80,
0.82, 0.85, 0.90, 0.95, 1.0 GeV. 10000 events were simulated for each
energy point. Simulated events were processed by the standard
offline reconstruction program and the standard selection criteria for
collinear events were applied. Then the energy deposition of the simulated
muons was corrected as described in sec.\ \ref{ecorsec}.

The resulting energy deposition for each energy point was parametrized
by the following function: 
\begin{eqnarray*}
\lefteqn{
  f^\mu_{SIM}(E) = p_1 \cdot \left[ \, 
p_2 \cdot g_l(E; E_0, \sigma_0, \eta) + \right. }
\qquad & & \\
& & \left. +(1-p_2)\cdot g_l(E; E_0,
\beta\sigma_0, \gamma\eta)
\right] + (1-p_1)\cdot g(E; E_1, \sigma_1) ,
\end{eqnarray*}
where $p_1$, $p_2$, $E_0$, $\sigma_0$, $E_1$, $\sigma_1$, $\eta$,
$\beta$ and $\gamma$ are the free fit parameters.

In the simulation we have a perfectly calibrated calorimeter while
the real data have an additional spread due to the uncertainty of the
calibration coefficients. As a result, the real energy distribution is
wider than that in the simulation. It could also be slightly
shifted. To take that into account, the energy deposition distribution
obtained from the simulation was modified to reflect the energy
deposition of real muons: 

\[
\begin{array}{c}
f_{\mu }^{+}(E)\equiv f_{\mu }^{-}(E)\equiv f_{\mu }(E), 
\vphantom{\Big /} \\
f_\mu (E) = f^\mu_{MIP}(E;k,\sigma_x),
\vphantom{\bigg /} \\
\multicolumn{1}{l}{
f^\mu_{MIP}(E; k, \sigma_x) = p_1 \cdot \left[ \, 
p_2 \cdot g_l(\tilde{E}; E_0, \tilde{\sigma}_0, \eta) + 
\right.
\vphantom{\bigg /} } \\
\multicolumn{1}{r}{
\qquad \qquad \qquad \left.
+(1-p_2)\cdot g_l(\tilde{E}; E_0, \beta\tilde{\sigma}_0, \gamma\eta)
\right] + (1-p_1)\cdot g(\tilde{E}; E_1, \tilde{\sigma_1}),
\vphantom{\bigg /} } \\
\tilde{E}=E/k, 
\quad \tilde{\sigma}_0=\sqrt{\sigma_0^2+\sigma_x^2}, 
\quad \tilde{\sigma}_1=\sqrt{\sigma_1^2+\sigma_x^2}.
\vphantom{\Big /}
\end{array}
\]

The values of $p_1$, $p_2$, $E_0$, $\sigma_0$, $E_1$, $\sigma_1$, $\eta$,
$\beta$ and $\gamma$ are fixed according to the simulation, so 
this function has only two free parameters: $k$ is the scale
coefficient and $\sigma_x$ is the additional energy spread. 

An example of the
fit for one energy point as well as the energy dependence of $E_0$ and
$\sigma_0$ parameters are shown in Fig.\ \ref{mip_mu_sim}. The $k$ and
$\sigma_x$ parameters for all energy points are shown in
Fig.~\ref{pipar}a and \ref{pipar}b respectively.

\vspace{2mm}

\subsubsection{Energy deposition of pions}

Pions produced in the reaction \( e^{+}e^{-}\rightarrow \pi ^{+}\pi
^{-} \) may 
interact with the calorimeter in two ways: as minimum ionising
particles or through nuclear interaction. 

\begin{figure}
\begin{center}
\begin{tabular}{cc}
\multicolumn{2}{c}
{\subfigure[Energy deposition of 400 MeV pions]
{\includegraphics[width=0.95\textwidth]{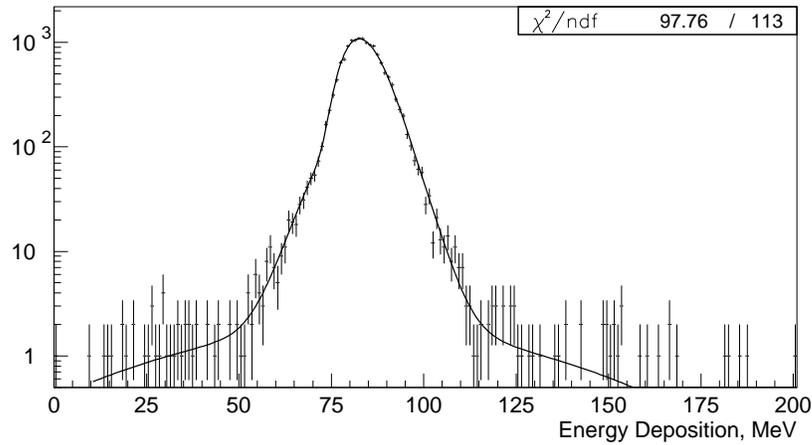}}} \\
\subfigure[Energy dependence of $E_0$]
{\includegraphics[width=0.46\textwidth]{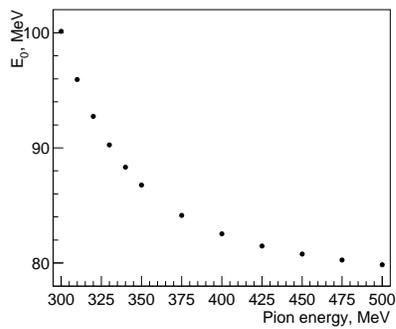}} &
\subfigure[Energy dependence of $\sigma_0$]
{\includegraphics[width=0.46\textwidth]{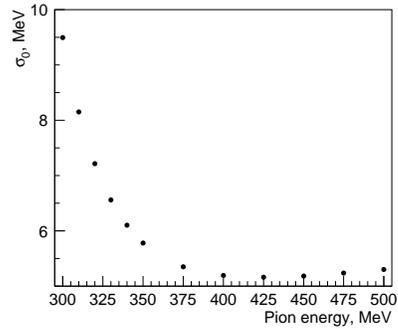}} \\
\end{tabular}
\end{center}
\caption{\label{mip_pi_sim} Simulated energy deposition of
minimum ionising pions}
\end{figure}

The energy deposition of the minimum ionising pions was obtained from
the detector simulation in the same way as for
muons. The simulation was performed for the center-of-mass energies 
0.6, 0.62, 0.64, 0.66, 0.68, 0.70, 0.75, 0.80, 0.85, 0.90, 0.95, 1.0 GeV.
10000 events were simulated for each
energy point. Simulated events were processed by the standard
offline reconstruction program and the standard selection criteria for
collinear events were applied. Then the energy deposition of the simulated
pions was corrected as described in sec.\ \ref{ecorsec}.

The resulting energy deposition for each energy point was parametrized
by the same function as for muons: 
\begin{eqnarray*}
\lefteqn{
  f^\pi_{SIM}(E) = p_1 \cdot \left[ \, 
p_2 \cdot g_l(E; E_0, \sigma_0, \eta) + \right. }
\qquad & & \\
& & \left. +(1-p_2)\cdot g_l(E; E_0,
\beta\sigma_0, \gamma\eta)
\right] + (1-p_1)\cdot g(E; E_1, \sigma_1) ,
\end{eqnarray*}
where $p_1$, $p_2$, $E_0$, $\sigma_0$, $E_1$, $\sigma_1$, $\eta$,
$\beta$ and $\gamma$ are the free fit parameters depending on the beam
energy.  This function describes well the energy deposition only for those
pions that do not stop in the calorimeter.

As was done for muons, the modified simulated energy
deposition was used for the parametrization of the energy deposition
of real pions:
\[
\begin{array}{c}
\multicolumn{1}{l}{
f^\pi_{MIP}(E; k, \sigma_x) = p_1 \cdot \left[ \, 
p_2 \cdot g_l(\tilde{E}; E_0, \tilde{\sigma}_0, \eta) + \right.
\vphantom{\bigg /} } \\
\multicolumn{1}{r}{
\qquad \qquad \qquad \left.
+ (1-p_2)\cdot g_l(\tilde{E}; E_0, \beta\tilde{\sigma}_0, \gamma\eta)
\right] + (1-p_1)\cdot g(\tilde{E}; E_1, \tilde{\sigma_1}),
\vphantom{\bigg /} } \\
\tilde{E}=E/k, 
\quad \tilde{\sigma}_0=\sqrt{\sigma_0^2+\sigma_x^2}, 
\quad \tilde{\sigma}_1=\sqrt{\sigma_1^2+\sigma_x^2}. 
\vphantom{\Big /}
\end{array}
\]
The values of $p_1$, $p_2$, $E_0$, $\sigma_0$, $E_1$, $\sigma_1$, $\eta$,
$\beta$ and $\gamma$ were fixed according to the simulation, so 
this function has only two free parameters: the scale
coefficient $k$ and the additional energy spread $\sigma_x$. 

An example of the energy deposition of minimum ionising pions and the
energy dependence of $E_0$ and 
$\sigma_0$ parameters are shown in Fig.\ \ref{mip_pi_sim}.
The $k$ and
$\sigma_x$ parameters for all energy points are shown in
Fig.~\ref{pipar}a and \ref{pipar}b respectively.

The energy deposition of the nuclear interacted pions cannot be
simulated well. Therefore, in this case we have chosen to use an empirical
parametrization, which fits experimental data well:
\[
f^\pi_{NI} (E; \sigma, p_{1\ldots N}) = \frac{g(E; 0,\sigma)+
\sum\limits_{i=1}^{N} g(E; \frac{E_0}{N+1}i, \sigma)\cdot p_i^2}{
1+\sum\limits_{i=1}^{N} p_i^2},
\]
where $E_0$ is the beam energy, N is the number of Gaussian and $\sigma$
and $p_{1\ldots N}$ are the free fit parameters. All Gaussian are
normalized at the $[0,+\infty]$, therefore $f^\pi_{NI}$ is also
normalized. This function appears as a set of equidistant Gaussians
of the same width but different area. The parametrization with $N=5$ or
$6$ describes well the experimental data.

The energy depositions of the minimum ionising $\pi^+$ and $\pi^-$ are
the same while they are quite different for nuclear interacted $\pi^+$ and
$\pi^-$. Unlike electrons and muons, pions have a significant
probability to have no cluster in the calorimeter and this
probability is different for $\pi^+$ and $\pi^-$. According to this,
the overall pion energy deposition was parametrized with the following
functions:
\begin{eqnarray}
\label{fpipip}
\lefteqn{ 
f_{\pi }^{+}(E) = p_{0}^{+}\cdot\delta(E)+(1-p_{0}^{+})\times
} & & \\
\nonumber
& & \times \left[ \, p^{+}_{MIP}\cdot f^\pi_{MIP}(E; k, \sigma_x)
+(1-p^{+}_{MIP})\cdot f^\pi_{NI}(E;\sigma^+, p^+_{1\ldots N})\right], 
\vphantom{\bigg /} \\
\label{fpipim}
\lefteqn{ 
f_{\pi }^{-}(E)  = p_{0}^{-}\cdot\delta(E)+(1-p_{0}^{-})\times
} \\
\nonumber
& & \times \left[ \, p^{-}_{MIP}\cdot f^\pi_{MIP}(E; k, \sigma_x)
+(1-p^{-}_{MIP})\cdot f^\pi_{NI}(E;\sigma^-, p^-_{1\ldots N})\right],
\vphantom{\bigg /} 
\end{eqnarray}
where $p_0^\pm$, $p_{MIP}^\pm$, $k$, $\sigma_x$, $\sigma^\pm$,
$p^\pm_{1\ldots N}$ are free fit parameters.

\begin{figure}
\begin{center}
\begin{tabular}{cc}
\subfigure[$\pi^+$, $345<E_\pi<355$ MeV]
%$\varphi\rightarrow\pi^+\pi^-\pi^0$ data]
{\includegraphics[width=0.45\textwidth]{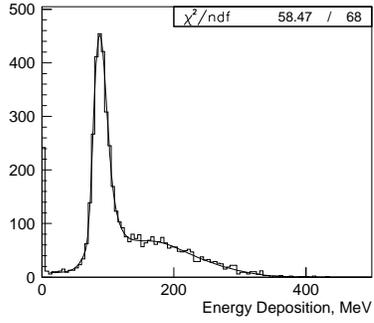}} &
\subfigure[$\pi^-$, $345<E_\pi<355$ MeV]
%$\varphi\rightarrow\pi^+\pi^-\pi^0$ data]
{\includegraphics[width=0.45\textwidth]{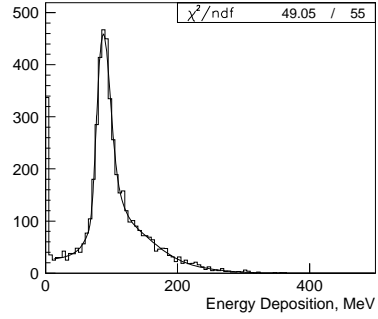}} \\
\subfigure[$\pi^+$, $425<E_\pi<435$ MeV]
%$\varphi\rightarrow\pi^+\pi^-\pi^0$ data]
{\includegraphics[width=0.45\textwidth]{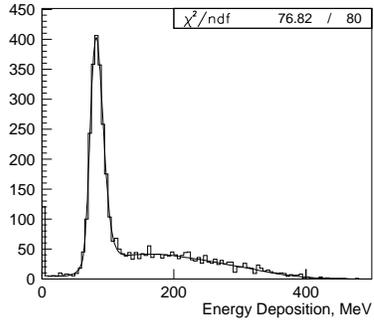}} &
\subfigure[$\pi^-$, $425<E_\pi<435$ MeV]
%$\varphi\rightarrow\pi^+\pi^-\pi^0$ data]
{\includegraphics[width=0.45\textwidth]{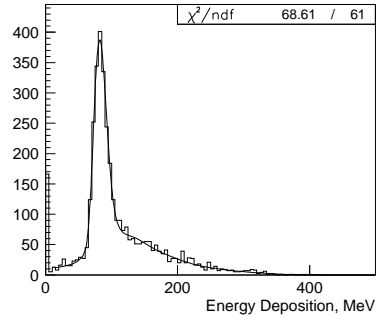}} 
\end{tabular}
\end{center}
\caption{\label{piedepex}An example of the experimental energy
deposition for pions of different energy. Pions were extracted from
$\varphi\rightarrow\pi^+\pi^-\pi^0$ data. 
The energy depositions of $\pi^+$  and $\pi^-$ were fitted with
functions (\ref{fpipip}) and (\ref{fpipim}) respectively}
\end{figure}

Pions produced in the $\phi(1020)$ decay to $3\pi$ were used to test
whether the functions (\ref{fpipip}) and (\ref{fpipim}) describe the
energy deposition of the real pions well enough. The CMD-2 has
collected about 20 
pb$^{-1}$ in the $\varphi$-meson energy range. About 100000 ``clean''
$\pi^+$ and $\pi^-$ were selected from the completely reconstructed
$\varphi\rightarrow\pi^+\pi^-\pi^0$ events. The energy deposition of
$\pi^+$ ($\pi^-$) with energies $345<E_\pi<355$ MeV and
$425<E_\pi<435$ MeV together with the fit are shown in Fig.\
\ref{piedepex}a and \ref{piedepex}c (\ref{piedepex}b and
\ref{piedepex}d). 

\subsubsection{Fit results}

The minimization of the likelihood function (\ref{lfunc1}) was carried
out for all
energy points with the typical number of 10000 events per energy point with the
following 27 free fit parameters:

\begin{itemize}
\item global parameters
\[
(N_{ee}+N_{\mu \mu }),\quad \frac{N_{\pi \pi }}{N_{ee}+N_{\mu \mu }}
\]

\item Electron energy deposition parameters
\[
p_{W}, \quad E_{0}, \quad \sigma, \quad \eta \]

\item Muon energy deposition parameters
\[
k_{MIP}, \quad \sigma_x^\mu \]

\item Pion energy deposition parameters
\begin{eqnarray*}
& \sigma_x^\pi, \quad p_{0}^{+}, \quad p_{0}^{-}, 
\quad p_{MIP}^{+}, \quad p_{MIP}^{-}, & \\
& \sigma _{NI}^{+},\quad \sigma _{NI}^{-}, 
\quad p^+_{1\ldots N}, \quad p^-_{1\ldots N}, \quad N=6 &
\end{eqnarray*}

\end{itemize}

The number of background events was fixed during the fit. 
The scale parameter for minimum ionising muons and pions was the same
(listed about as $k_{MIP}$). 

\begin{figure}
\begin{center}
\begin{tabular}{cc}
\subfigure[$N_{ee}+N_{\mu\mu}$]
{\includegraphics[width=0.45\textwidth]{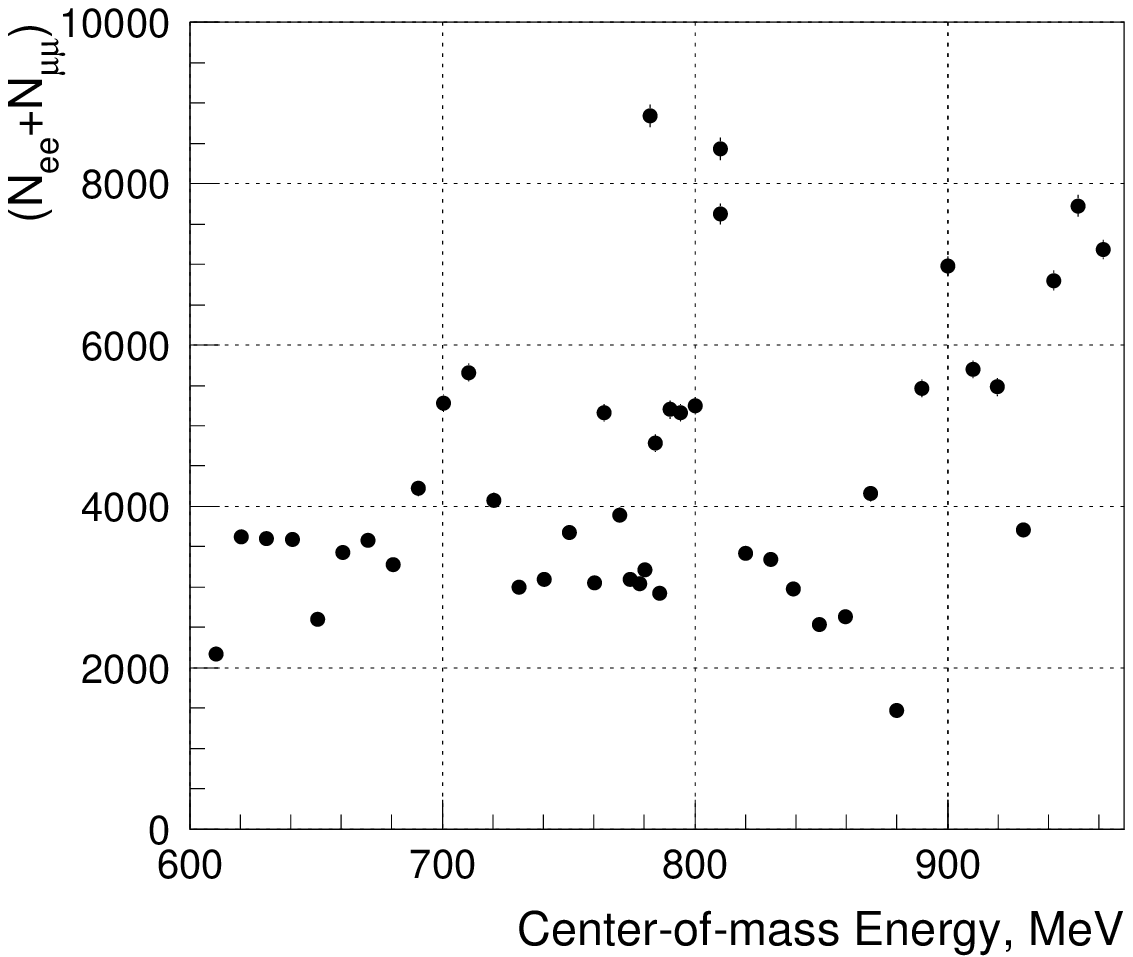}} &
\subfigure[$N_{\pi\pi}/(N_{ee}+N_{\mu\mu})$]
{\includegraphics[width=0.45\textwidth]{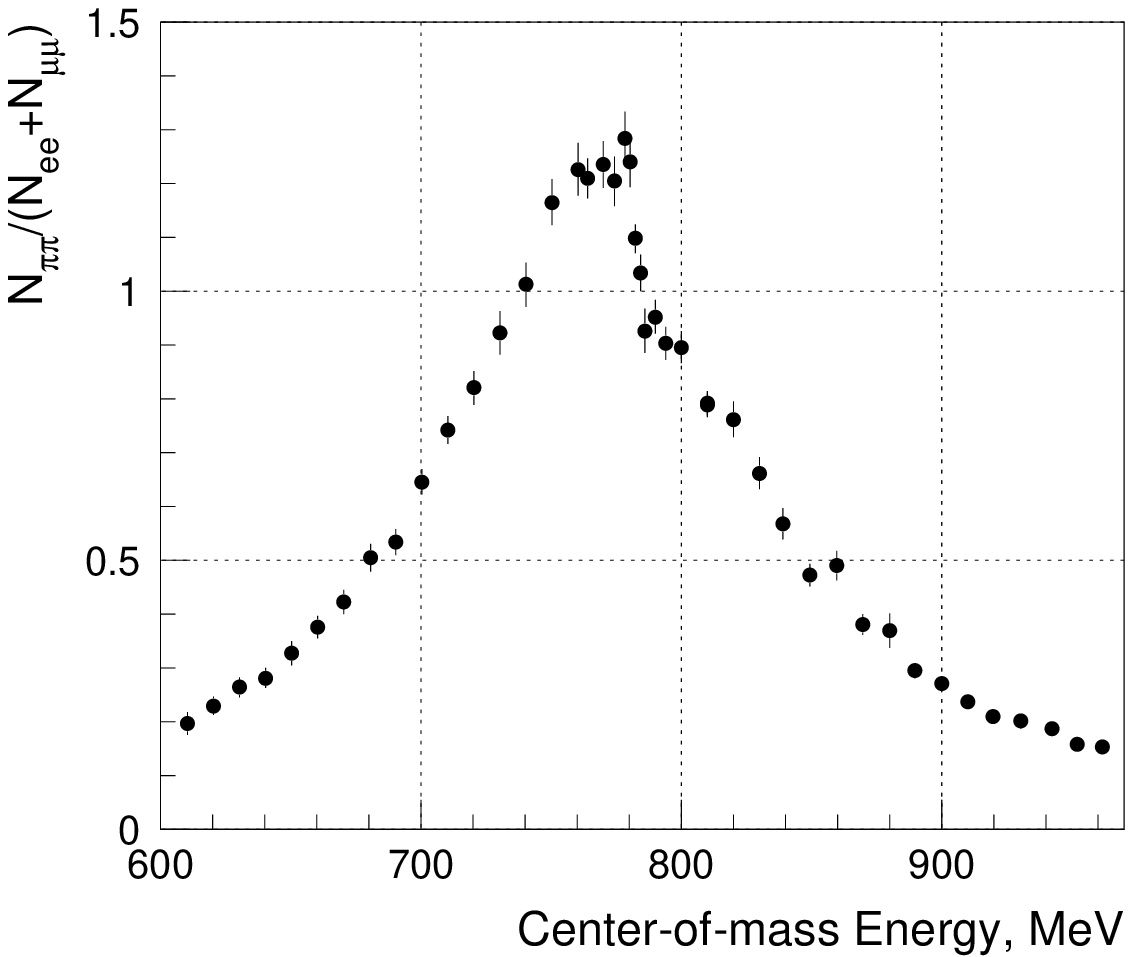}} 
\end{tabular}
\end{center}
\caption{\label{gpar} Results of the 
minimization of the likelihood function (\ref{lfunc1}).
The energy dependences of the global parameters are shown}
\end{figure}

\begin{figure}
\begin{center}
\begin{tabular}{cc}
\subfigure[$E_{0}$]
{\includegraphics[width=0.45\textwidth]{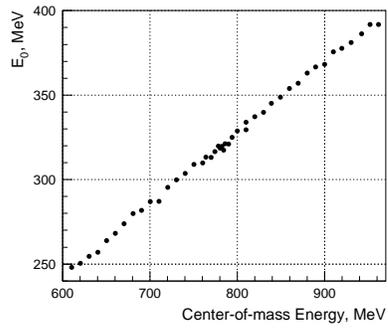}} &
\subfigure[$\sigma$]
{\includegraphics[width=0.45\textwidth]{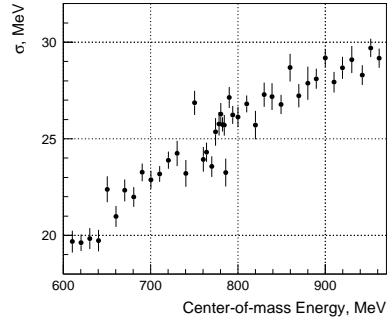}} \\
\subfigure[$p_W$]
{\includegraphics[width=0.45\textwidth]{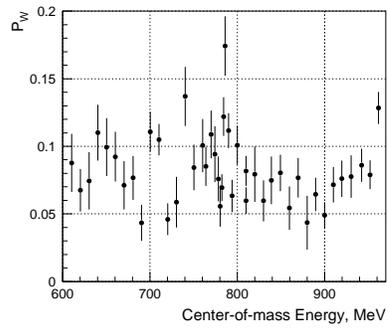}} &
\subfigure[$\eta$]
{\includegraphics[width=0.45\textwidth]{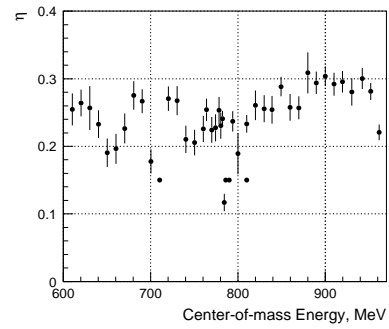}}
\end{tabular}
\end{center}
\caption{\label{epar} Results of the 
minimization of the likelihood function (\ref{lfunc1}).
The electron energy deposition parameters are shown}
\end{figure}

\begin{figure}
\begin{center}
\begin{tabular}{cc}
\subfigure[$k_{MIP}$]
{\includegraphics[width=0.45\textwidth]{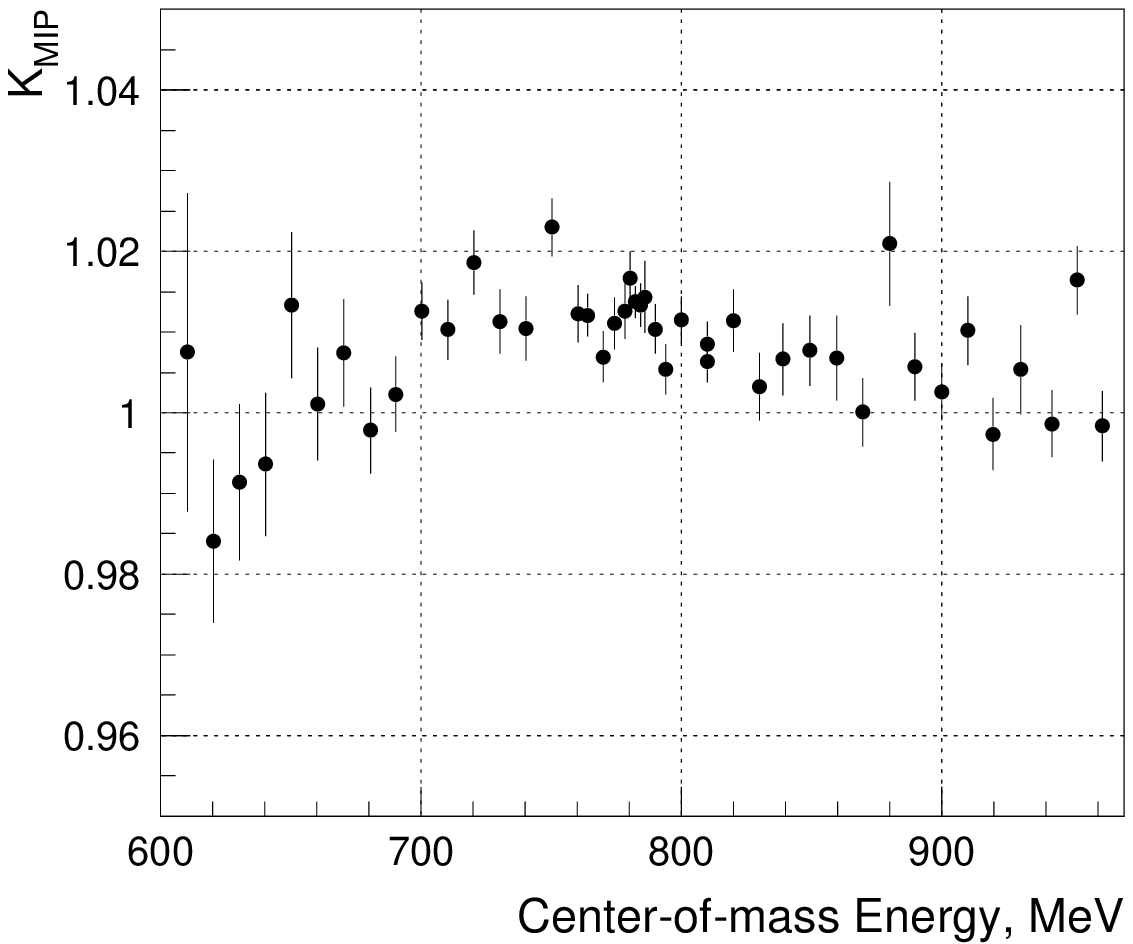}} &
\subfigure[$\sigma_x^\mu$ and $\sigma_x^\pi$]
{\includegraphics[width=0.45\textwidth]{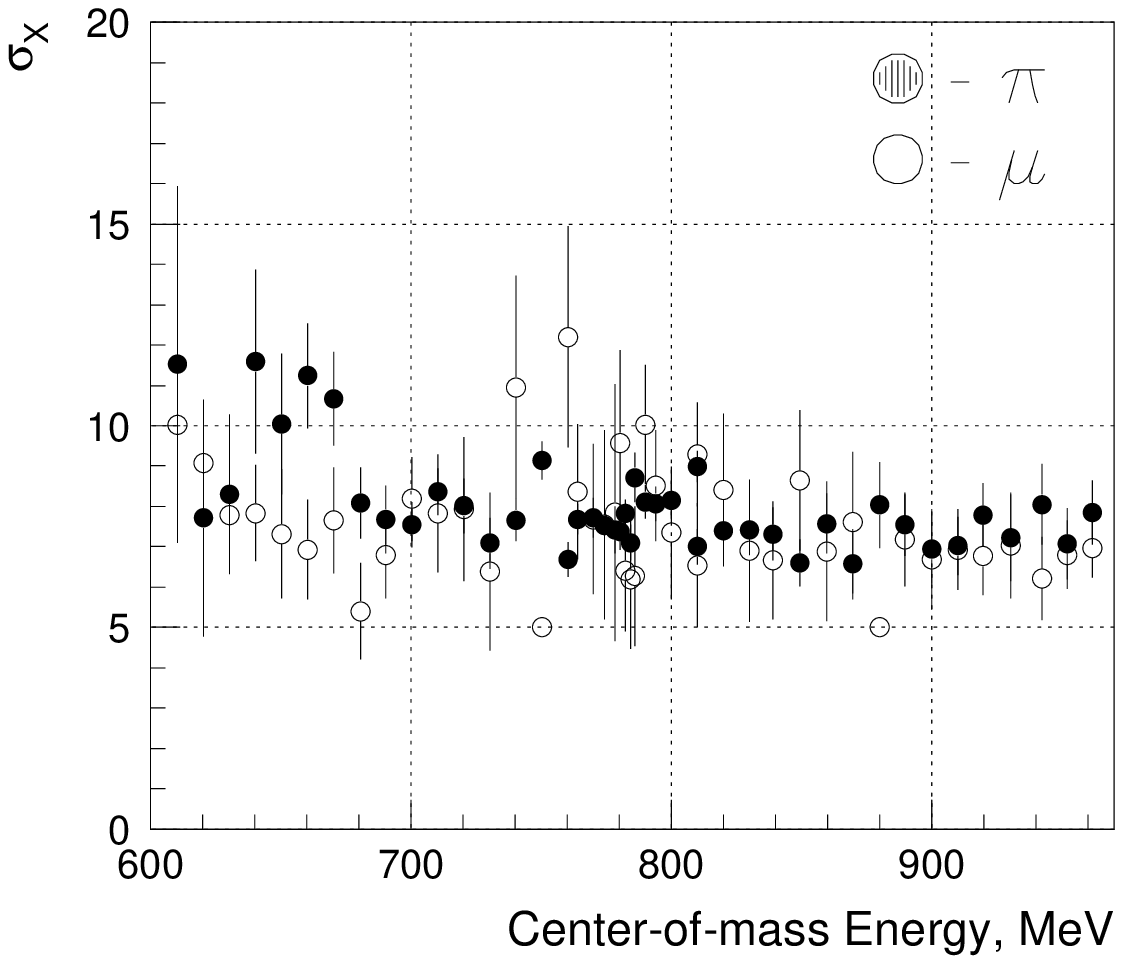}} \\
\subfigure[$p_0^+$ and $p_0^-$]
{\includegraphics[width=0.45\textwidth]{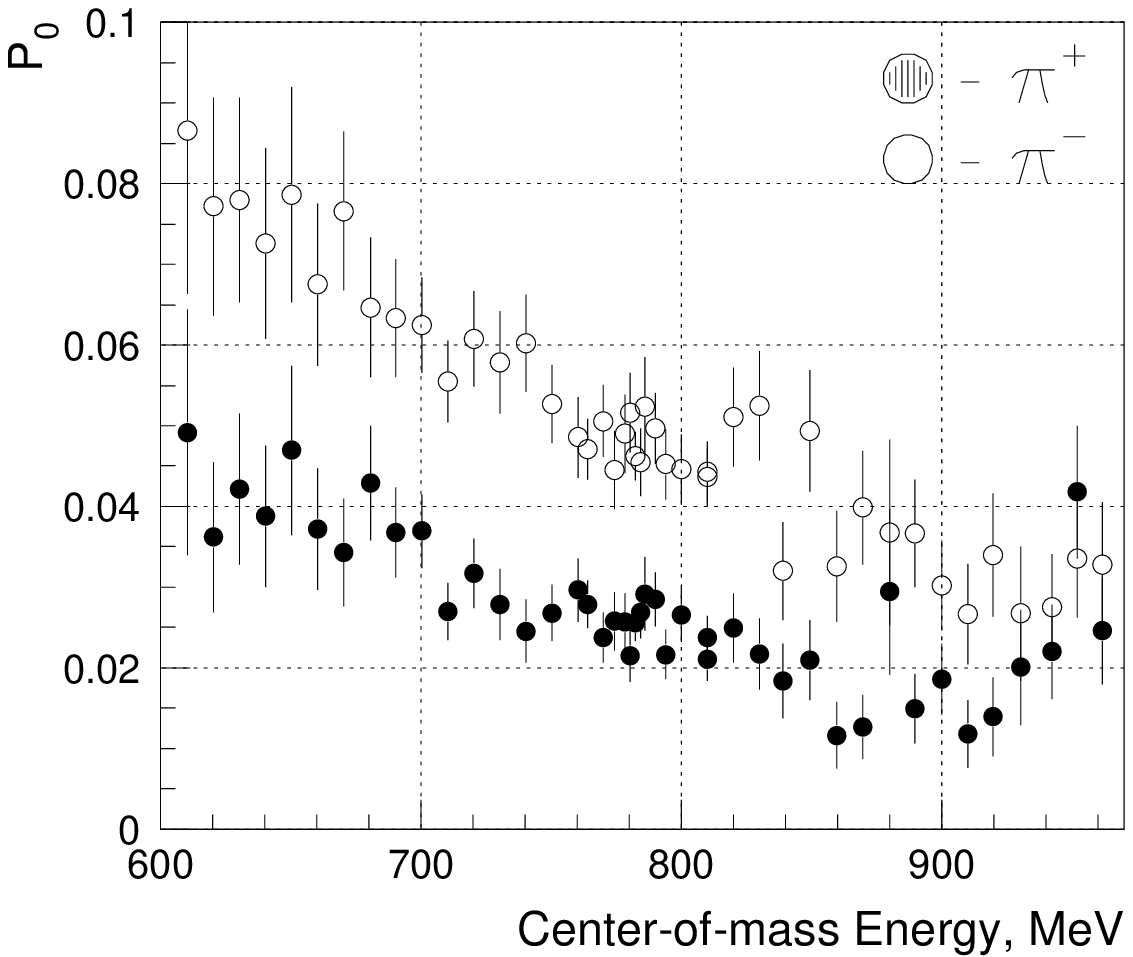}} &
\subfigure[$p_{MIP}^+$ and $p_{MIP}^-$]
{\includegraphics[width=0.45\textwidth]{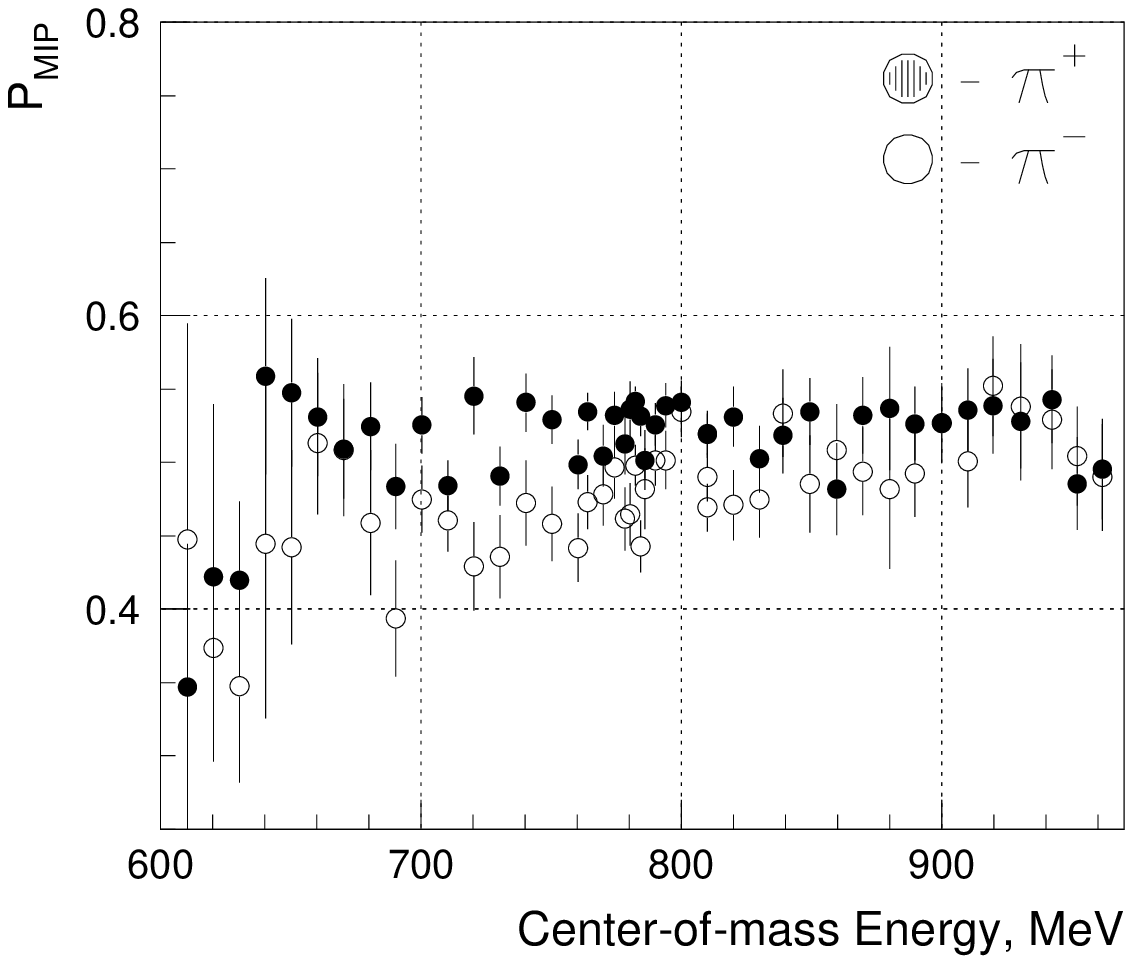}} 
\end{tabular}
\end{center}
\caption{\label{pipar} Results of the 
minimization of the likelihood function (\ref{lfunc1}).
The muon and pion energy deposition parameters are shown}
\end{figure}

The results of the minimization are shown in
Fig.~\ref{gpar}--\ref{pipar}: the global parameters are shown in
Fig.~\ref{gpar}, the electron energy deposition parameters are shown
in Fig.~\ref{epar} and the muon and pion energy deposition parameters
are shown in Fig.~\ref{pipar}. 

\newpage

\subsection{\label{piformsec}Form factor calculation}

After the minimization of the likelihood function, the pion form
factor was calculated as:
\begin{eqnarray}
\label{piform}
\lefteqn{
\left| F_{\pi }\right| ^{2}=
\frac{N_{\pi \pi }}{N_{ee}+N_{\mu \mu }}\times
} & & \\
\nonumber & & \times \frac{\sigma ^{B}_{ee}\cdot \left( 1+\delta _{ee}\right)
\left( 1+\alpha _{ee}\right) \varepsilon _{ee}+
\sigma ^{B}_{\mu \mu }\cdot \left( 1+\delta _{\mu \mu }\right)
\left( 1+\alpha _{\mu \mu }\right) \varepsilon _{\mu \mu }}
{\sigma ^{B}_{\pi \pi }\cdot \left( 1+\delta _{\pi \pi }\right)
\left( 1+\alpha _{\pi \pi }\right)
\left( 1-\Delta _{H}\right) \left( 1-\Delta _{D}\right)
\varepsilon _{\pi \pi }}- \\
\nonumber & & - \Delta _{3\pi },
\end{eqnarray}
where \( N_{\pi \pi }/(N_{ee}+N_{\mu \mu }) \) was obtained as a result of
minimization, \( \sigma ^{B}_{a} \) are the Born cross-sections, \( \delta _{a} \)
are the radiative corrections, \( \varepsilon _{a} \) are the efficiencies, including
trigger and reconstruction efficiencies, \( \alpha _{a} \) are the corrections
for the experimental resolution of the \( \Theta  \) angle measurement, \( \Delta _{H} \)
is the correction for pion losses due to nuclear interactions, \( \Delta _{D} \)
is the correction for pion losses due to decays in flight and \( \Delta _{3\pi } \)
is the correction for the misidentification of \( \pi ^{+}\pi ^{-}\pi ^{0} \)
events as a \( \pi ^{+}\pi ^{-} \) pair.

Together with the pion form factor, the cross-section of $e^{+}e^{-}$
annihilation to $\pi^+\pi^-$ was calculated as 
\[
\sigma _{\pi \pi }=\frac{\pi \alpha ^{2}}{3s}\left( 1-\frac{4m_{\pi
}^{2}}{s}\right) ^{3/2}\cdot \left| F_{\pi }\right| ^{2},
\]
 and the luminosity integral as
\[
\int Ldt=\frac{N_{ee}+N_{\mu \mu }}{\sigma _{ee}^{B}\cdot (1+\delta _{ee})(1+\alpha _{ee})\varepsilon _{ee}+\sigma _{\mu \mu }^{B}\cdot (1+\delta _{\mu \mu })(1+\alpha _{\mu \mu })\varepsilon _{\mu \mu }}.\]

\subsubsection{Radiative corrections}

\begin{figure}[tb]
\begin{center}
\includegraphics[width=0.9\textwidth]{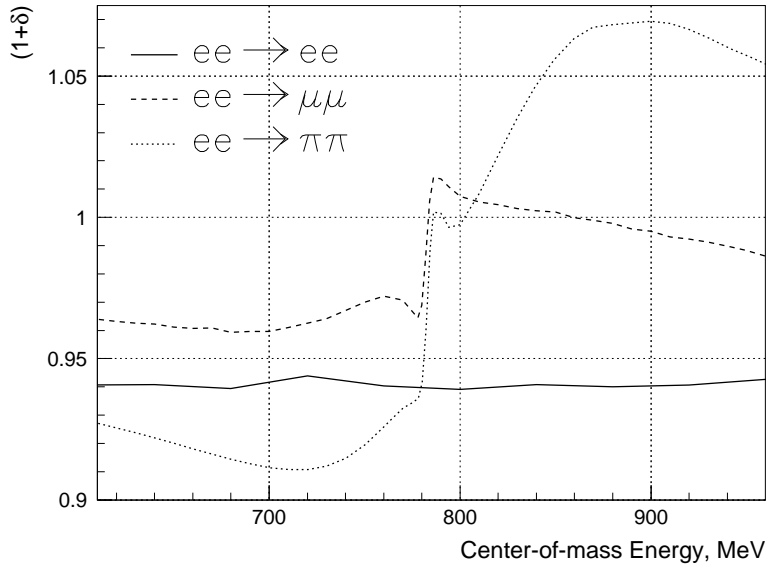}
\end{center}
\caption{\label{rcorfig}
Radiative corrections for $e^+e^-\rightarrow e^+e^-, \pi^+\pi^-,
\mu^+\mu^-$ events}
\end{figure} 

The radiative corrections are shown in Fig.~\ref{rcorfig}.
The calculation of the radiative corrections for $e^+e^-\rightarrow e^+e^-$
events was based on \cite{radee}. The estimated systematic error is
$\sim 1\%$. Radiative
corrections for \( e^{+}e^{-}\rightarrow \mu ^{+}\mu ^{-},\pi ^{+}\pi ^{-} \)
were calculated according to \cite{rc1} and \cite{rc2}. The estimated
systematic error is $0.2-0.5\%$.
The contribution from the lepton and hadron vacuum polarization
is included in the radiative corrections for \( e^{+}e^{-}\rightarrow
e^{+}e^{-},\mu ^{+}\mu ^{-} \), 
but excluded from the radiative correction for 
\( e^{+}e^{-}\rightarrow \pi ^{+}\pi ^{-} \). 
Some details about the radiative correction calculation could be found
in \cite{PREP}.

The radiative correction for $e^+e^-\rightarrow \pi^+\pi^-$ 
depends on the energy behaviour of the $e^+e^-\rightarrow \pi^+\pi^-$
cross-section itself. In order to take that into account, the
calculation of this correction was done via a few iterations. At the
first iteration, the existing $\left|F_\pi(s)\right|^2$ data were
used for the calculation of the radiative correction. Using the calculated
correction, we obtain $\left|F_\pi(s)\right|^2$. At subsequent
iterations, $\left|F_\pi(s)\right|^2$ data obtained from the previous
step, were used. It was found that after 3 iterations
$\left|F_\pi(s)\right|^2$ values become stable.

\subsubsection{Trigger efficiency}

The trackfinder (TF) signal, neutral trigger (NT) signal and the calorimeter
``OR'' (CSI) trigger signal have been used for detector triggering during
1994-1995 runs. The trackfinder uses only the drift chamber and the
Z-chamber data for a fast decision
whether there is at least one track in the event. The neutral trigger
analyses the geometry 
and energy deposition of the fired calorimeter rows. The calorimeter ``OR''
signal appears when there is at least one triggered calorimeter row. 

Two different trigger settings have been used during 1994-1995 runs. 
For 810(2)-960
MeV energy points the trigger was ``\textbf{(TF.and.CSI).or.NT}''
while for other energy points the trigger was
``\textbf{TF.or.NT}''. 
For analysis we use only those events where the trackfinder has found
the track. If the trackfinder efficiency is $\varepsilon_{TF}$ and the
CSI signal efficiency is $\varepsilon_{CSI}$, the overall trigger
efficiency is $\varepsilon_{TF}\cdot\varepsilon_{CSI}$ for 810(2)-960
MeV energy points and $\varepsilon_{TF}$ for 610-810(1) MeV energy
points. 

\begin{figure}
\begin{center}
\begin{tabular}{c}
\subfigure[\label{tfefffig}
Trackfinder efficiency for $e^+e^-\rightarrow e^+e^-$ events]
{\includegraphics[width=0.75\textwidth]{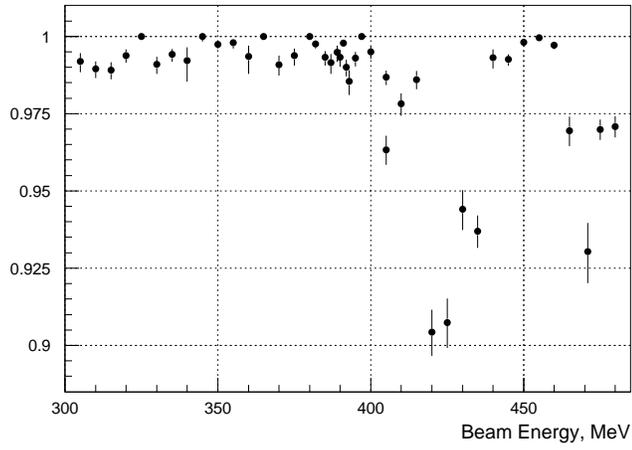}} \\
\subfigure[\label{csieff}CSI signal efficiency for collinear events]
{\includegraphics[width=0.75\textwidth]{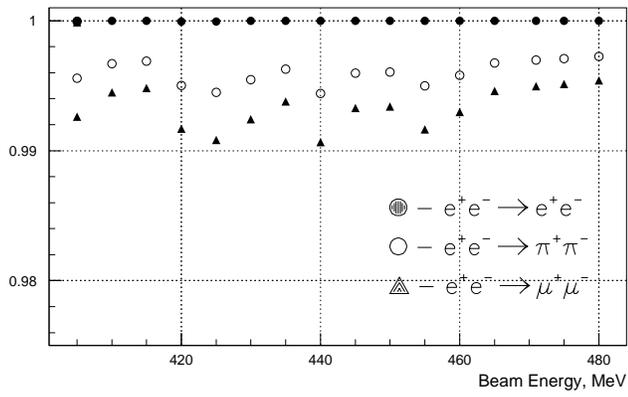}}
\end{tabular}
\end{center}
\caption{\label{trgeff}Trigger efficiency for collinear events}
\end{figure}

For the measurement of the trackfinder efficiency $e^+e^-\rightarrow
e^+e^-$  events have been
selected using only the calorimeter data (see the next section for
details). Additionally, the positive decision of the neutral trigger
was required. Analysing the probability to have also a positive
decision of the trackfinder, its efficiency for $e^+e^-\rightarrow
e^+e^-$ events can be calculated. The result is shown in Fig.\ \ref{tfefffig}. 
All types of collinear events are similar from the trackfinder point of
view --- they have two collinear tracks and the track momenta are very
close for different types of events. It is seen from the
picture that the TF efficiency is high enough and that there is no visible
energy dependence of the efficiency. Therefore we assume that the
trackfinder efficiency is the same for all types of collinear events
and hence cancels in (\ref{piform}).

For the measurement of the CSI signal  efficiency the fact that there
are two clusters 
in the collinear event have been used: the CSI efficiency for a single cluster
has been measured and was used for the calculation of the CSI efficiency for
the event. The collinear events have been selected using the standard
cuts. Then, the probability to trigger the CSI signal was measured for one
cluster as a function of its energy. Using the measured probability and
our knowledge about the energy deposition of particles of different
types, the efficiency to trigger the CSI signal was calculated for 
different types of events. Results for 810(2)-960 MeV energy points
are shown in Fig.\ \ref{csieff}. 
For simplicity, the values of 99.5\% for pions and 99.2\%
for muons for the CSI efficiency were used for all 810-960 MeV energy points.

\subsubsection{\label{receffsec}Reconstruction efficiency}

\begin{figure}
{\centering \resizebox*{0.9\textwidth}{!}{\includegraphics{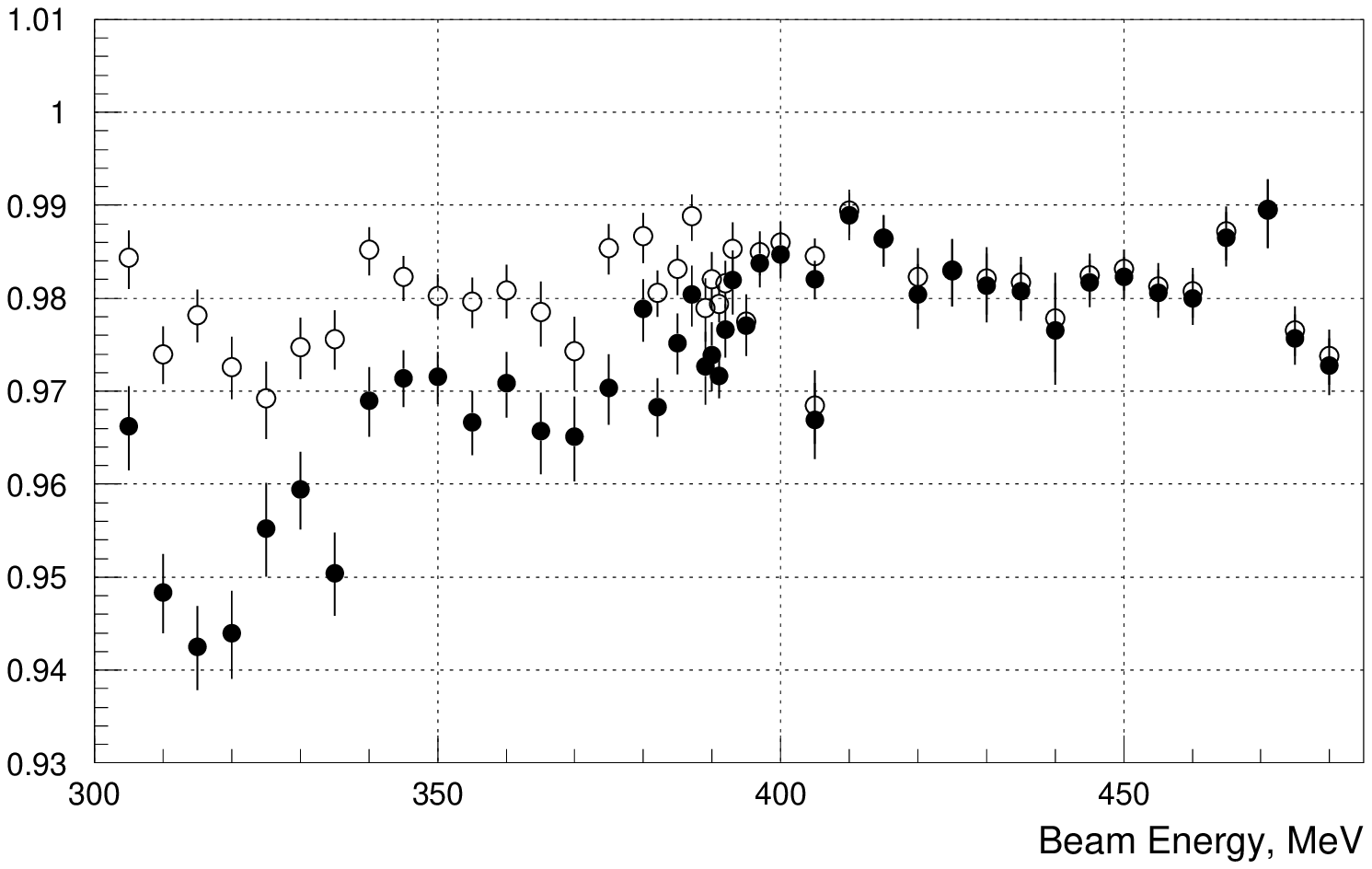}} \par}
\caption{\label{receff}Reconstruction efficiency for all energy points}
\end{figure}

The reconstruction efficiency was measured for $e^+e^-\rightarrow
e^+e^-$ events using the experimental data: events were selected using only
the calorimeter data, and then the probability to reconstruct two
collinear tracks was measured. Since the selection criteria for
collinear events are based on tracking data only, the measured
probability is the reconstruction efficiency. 

The following selection criteria based only on the calorimeter data,
were used.
\begin{enumerate}
\item There are exactly two clusters in the calorimeter.
\item There is a hit in the Z-chamber near each cluster. This
requirement selects the clusters produced by the charged particle.
\item The energy deposition of both clusters is between \(
(0.82\cdot E_{B}-40) \) and \( (0.82\cdot E_{B}+50) \) MeV.
This is typical for $e^+$ and $e^-$, but not for other particles.
\item The clusters are collinear if one takes into account the particle
motion in the detector magnetic field: 
\begin{eqnarray*}
& |\pi -(\Theta _{1}+\Theta _{2})|<0.1, & \\
& \left| \left| \pi -|\varphi _{1}-\varphi _{2}|\right| -2\arcsin 
\left( \frac{R\cdot 0.3B}{2\cdot E_{B} \cdot \sin \Theta }\right) \right|
<0.1, &
\end{eqnarray*}
where \( \Theta  \) and \( \varphi  \) are the polar and azimuthal
angles of the cluster, \( R \) is the effective calorimeter radius (45
cm) and \( B \) is the magnetic field (typically 10 kGs).
\item The event was triggered by the trackfinder. Other triggers may
be present.
\end{enumerate}
The selected sample contain only \( e^{+}e^{-}\rightarrow e^{+}e^{-} \) events
with negligible ($<0.5\%$) background. If the number of selected events
is \( N_{0} \) and the number of events, which in addition satisfy the
standard selection criteria for collinear events is $N_1$, 
then the reconstruction efficiency is 
\[
\varepsilon _{rec}={N_{1}}/{N_{0}}.\]
 The measured reconstruction efficiency is shown in Fig.\ \ref{receff} (filled
dots). The efficiency drops down for beam energies below 400
MeV. This additional inefficiency is completely explained by the lower
precision of the \( \Theta  \) 
angle measurement for these energy points. The empty dots in the same
plot represent 
the reconstruction efficiency for the same selection criteria for
collinear events  except for \( \Delta \Theta  \).

As in studying the trigger efficiency, all types of collinear events
are very similar from the point of view 
of the tracking reconstruction code. Therefore we assume that the
reconstruction efficiency  is the same for all types of
collinear events and hence it cancels in
(\ref{piform}). 

\subsubsection{Correction for the limited accuracy of the polar angle
measurement}

\begin{figure}
{\centering \begin{tabular}{cc}
\subfigure[$\Theta$-resolution]{\resizebox*{0.45\textwidth}{!}{\includegraphics{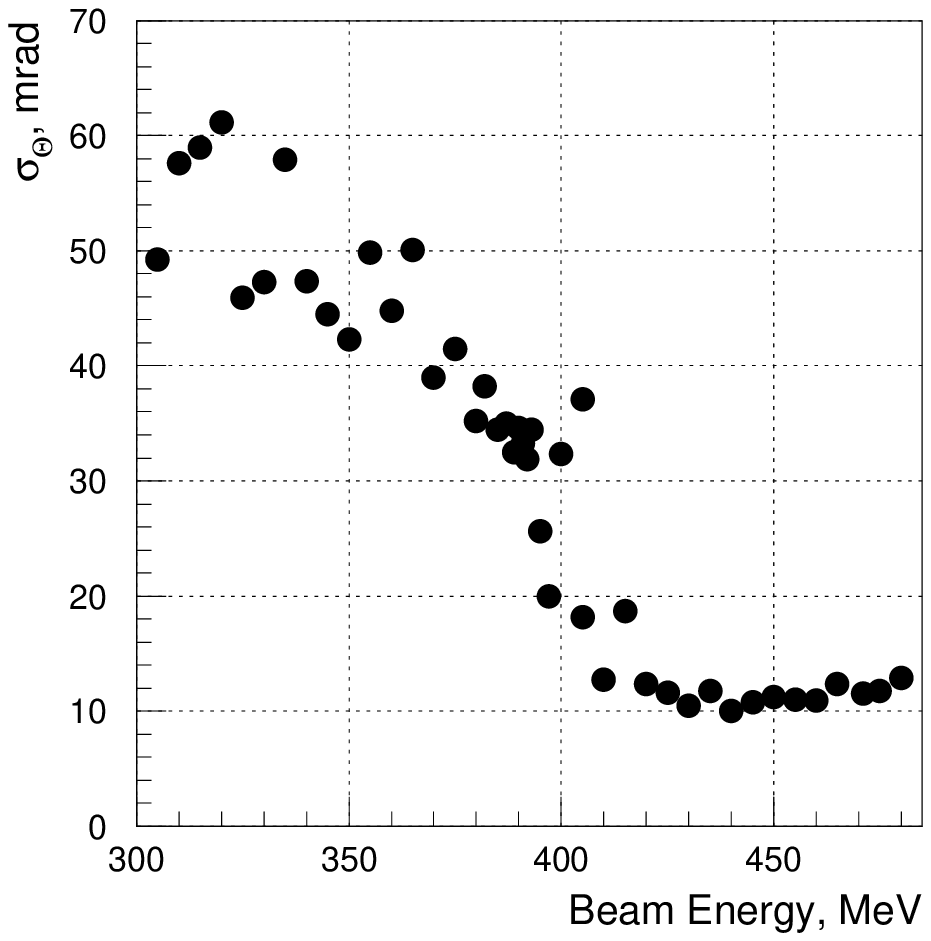}}} &
\subfigure[Correction to cross-section]{\resizebox*{0.45\textwidth}{!}{\includegraphics{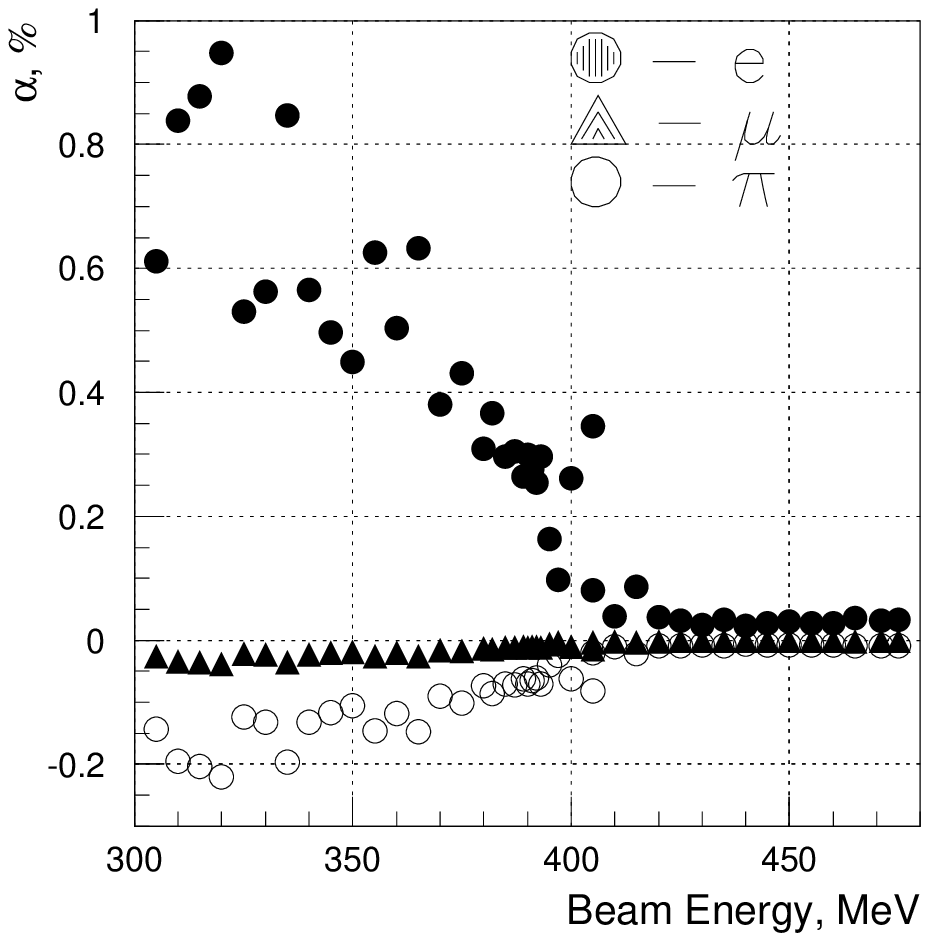}}} \\
\end{tabular}\par}
\caption{\label{dthcor}The experimental resolution of the average
polar angle measurement 
and the corresponding correction to cross-sections}
\end{figure}

There is a special type of correction that arises since there is finite
experimental resolution in the measurement of the $\Theta$ angles of
tracks. Since we require the measured, rather than real, 
$\;\Theta_{avr}=[\Theta_1+(\pi-\Theta_2)]/2\;$ to be
within $[\Theta_{min}, \pi-\Theta_{min}]$ interval, the visible
cross-section is slightly different from 
\[ \int_{\Theta_{min}}^{\pi-\Theta_{min}} \frac{d\sigma}{d\Omega} \cdot
2\pi\sin\Theta \, d\Theta. \] The corresponding correction, denoted
$\alpha$ in (\ref{piform}), was calculated as
\[
1+\alpha =\int\limits_{\Theta_{min}}^{\pi-\Theta_{min}}d\Theta 
\int\limits_{0}^{\pi}d\Theta^\prime \cdot f(\Theta^\prime) 
p(\Theta^\prime,\Theta) \left/ 
\int\limits_{\Theta_{min}}^{\pi-\Theta_{min}}d\Theta \cdot 
f(\Theta )\right. ,\]
where $f(\Theta)$ is the theoretical $\Theta$-distribution and 
$p(\Theta^\prime,\Theta)$ is the detector response function taken as
\[
p(\Theta ^{\prime },\Theta )=\frac{1}{\sqrt{2\pi }\sigma _{\Theta }}\exp \left[ -\frac{(\Theta ^{\prime }-\Theta )^{2}}{2\sigma _{\Theta }^{2}}\right] \]
where \( \sigma _{\Theta } \) is the experimental resolution of 
 \( \Theta_{avr} \) measurement. 
The resolution and the corresponding corrections for 
different types of collinear events for all energy points are
shown in Fig.\ \ref{dthcor}.

\subsubsection{Correction for pion losses due to nuclear interactions}

\begin{figure}
\begin{center}
\includegraphics[width=0.8\textwidth]{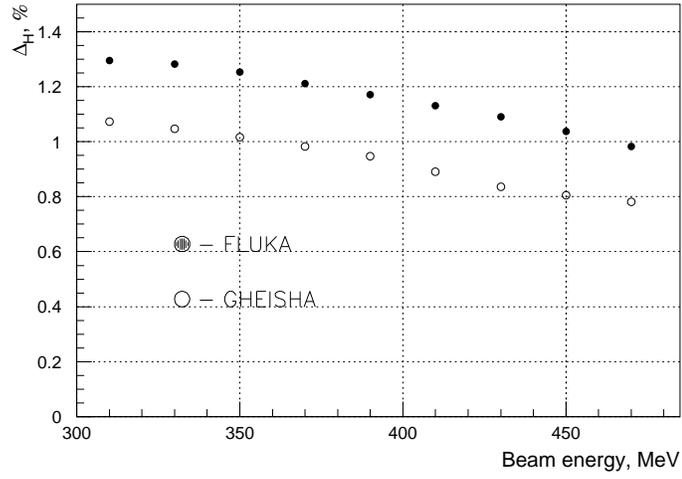}
\end{center}
\caption{\label{nuccorr}The probability $\Delta_H$ to lose 
the \protect\( e^{+}e^{-}\rightarrow \pi ^{+}\pi ^{-}\protect \) event from
the sample of collinear events due to nuclear interactions of pions
with the beam pipe or drift chamber inner support}
\end{figure}

Unlike electrons and muons, a small fraction of pions has nuclear
interactions in the beam pipe and therefore is lost from the selected
samples of collinear events. The corresponding
correction was calculated under the assumption that the event is lost
from the collinear event sample if at least one pion had an inelastic
nuclear interaction in the beam pipe or drift chamber inner
support.
The calculations were done separately for cross-sections obtained from two packages
available in GEANT\cite{Geant} for simulation of nuclear interactions ---
FLUKA\cite{FLUKA} and GHEISHA\cite{GHEISHA}. The resulting correction is shown
in Fig.\ \ref{nuccorr}. It has been shown \cite{Krokovny} that FLUKA is more
successful in describing experimental data about nuclear interactions of pions
in our energy range, and therefore the results obtained from FLUKA
were used for 
the cross-section correction while the difference between two calculations
was added to the systematic error.

\subsubsection{Correction for pion losses due to decays in flight}

\begin{figure}
\begin{center}
\includegraphics[width=0.8\textwidth]{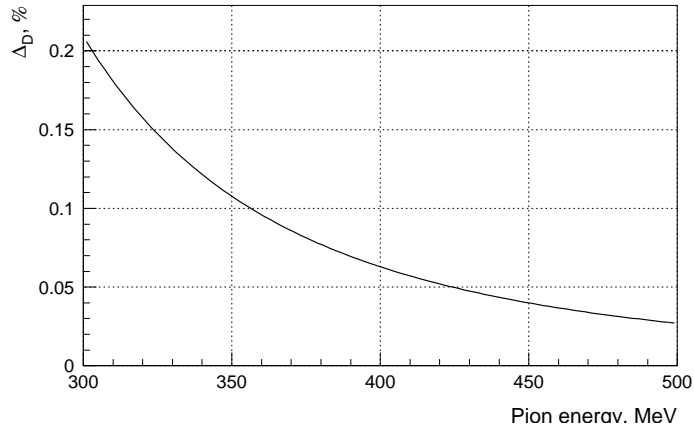}
\end{center}
\caption{\label{decaycorr}The probability $\Delta_D$ to lose 
the \protect\( e^{+}e^{-}\rightarrow \pi ^{+}\pi ^{-}\protect \) event from
the sample of collinear events due to pion decay in flight}
\end{figure}

A small fraction of pions decays in flight inside the drift chamber and the
corresponding event may not be recognized as collinear if the decay angle between
the pion and secondary muon is large enough. Depending on the beam energy
2-4\% of the $e^{+}e^{-}\rightarrow \pi ^{+}\pi ^{-}$ events have a decayed
pion. But the maximum decay angle for the analysed energy range is
small --- 80-150 
mrad depending on the pion energy --- and therefore more than 90\% of
such events 
are reconstructed as a pion pair. The probability to lose a pion pair
due to the pion 
decay in flight was found from simulation and is shown in Fig.~\ref{decaycorr}.

\subsubsection{Correction for the background from $\omega \rightarrow 3\pi$}

\begin{figure}
\begin{center}
\includegraphics[width=0.8\textwidth]{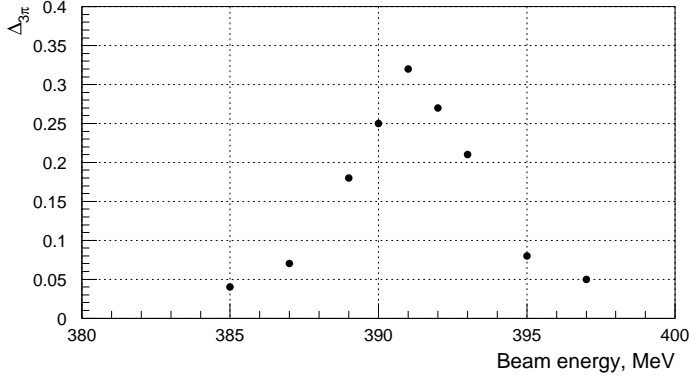}
\end{center}
\caption{\label{3picorr}The $\Delta_{3\pi}$ contribution to
the pion form factor from $\omega \rightarrow 3\pi$}
\end{figure}

There is a small probability to identify the $\pi^+\pi^-\pi^0$ final
state as a $\pi^+\pi^-$ pair. Therefore the measured pion form factor
is slightly bigger than it should 
be. This effect is significant only in the narrow energy range near the
$\omega$-meson. The corresponding probability was estimated from simulation and was found to
be \( 5\cdot 10^{-3} \) for \( \Theta _{min}=1.0 \) and \( 4\cdot 10^{-3} \)
for \( \Theta _{min}=1.1 \) rad. The contribution to the pion
form factor \( \Delta _{3\pi } \) was calculated assuming the  $\omega$-meson 
parameters from \cite{PDG98} and is presented in Fig.\ \ref{3picorr}.

\newpage

\subsection{Systematic errors}

The overall systematic error is estimated to be 1.5\% for 0.780
and 0.784 GeV, 1.7\% for 0.782, 0.84-0.87 and 0.94 GeV and 1.4\% for other
energy points. The main sources of systematic error are 
summarized in Table \ref{systerr} and discussed in detail below.

\begin{table}[h]
\begin{center}
\begin{tabular}{|c|c|}
\hline 
Source & Estimated value\\
\hline 
\hline 
Events separation & 0.6\% \\
\hline 
Energy calibration of collider & 0.1\% (0.5\% for 390,392; 1\% for 391)\\
\hline 
Fiducial volume & 0.5\% \\
\hline 
Trigger efficiency & 0.2\% (1\% for 420-435, 470) \\
\hline 
Reconstruction efficiency & 0.3\% \\
\hline 
Hadronic interactions of pions & 0.4\% \\
\hline 
Pions decays in flight & 0.1\% \\
\hline 
Radiative corrections & 1\% \\
\hline 
\hline 
Total & 1.4\% \\
& (1.5\% for 390,392)\\
& (1.7\% for 391,420-435,470)\\
\hline 
\end{tabular}
\end{center}
\caption{\label{systerr}Main sources of systematic errors}
\end{table} 

\paragraph{Event separation.}

\begin{figure}
\begin{center}
\begin{tabular}{c}
\subfigure[The relative difference between results of minimization of
two different likelihood functions]
{\includegraphics[height=0.3\textheight]{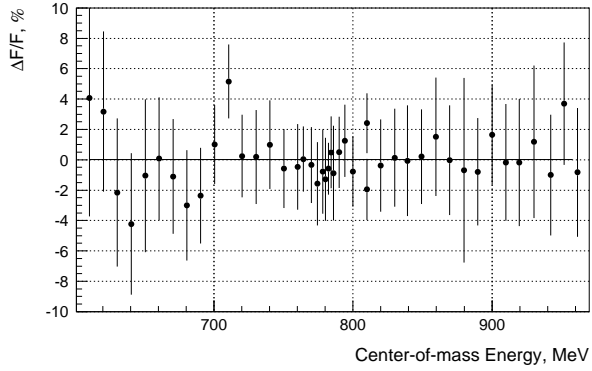}}  \\
\subfigure[The relative difference between simulated and reconstructed
form factor values for the complete detector simulation of collinear
events]
{\includegraphics[height=0.3\textheight]{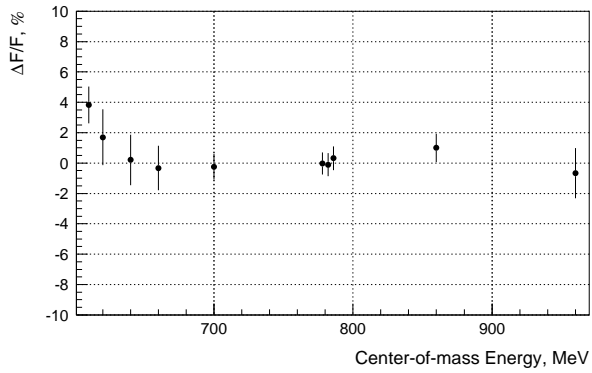}}  
%\\
%\subfigure[The relative difference between simulated and reconstructed
%form factor values for the simulation of different energy deposition
%of $e^+$ and $e^-$.]
%{\includegraphics[height=0.2\textheight]{fpisim3.eps}} 
\end{tabular}
\end{center}
\caption{\label{fpisim} Estimation of the systematic error of events
separation procedure. The relative difference $\Delta |F_\pi|^2 / |F_\pi|^2$ is
shown with the error bars representing statistical error}
\end{figure}

Special studies were performed to estimate the systematic error due to
event separation.
\begin{enumerate}
\item The separation was done by minimization of two independent
likelihood functions. The ``standard'' likelihood function was
described in detail in the previous sections. The second likelihood
function was different in the following ways: different 
separation of background events; different energy deposition functions
for minimum ionising muons and pions and for the nuclear interactions of
pions (a single Gaussian was used instead of the set of Gaussians).
The relative difference between two minimizations results is presented
in Fig.\ \ref{fpisim}a. The error bars represent the statistical
error. It is seen that there is no systematic shift and the
difference is well below the statistical error.
\item The complete detector simulation of $e^+e^-\rightarrow
e^+e^-(\gamma),\mu^+\mu^-,\pi^+\pi^-$ events in the GEANT environment was done
for 10 energy points: 0.61, 0.62, 0.64, 0.66, 0.70, 0.778, 0.782,
0.786, 0.860 and 0.960 GeV. 100000 of $e^+e^-\rightarrow
e^+e^-(\gamma)$ events and
the corresponding number of $e^+e^-\rightarrow\mu^+\mu^-,\pi^+\pi^-$
events were simulated for each energy point. The event separation was
done by minimization of the same likelihood function as for real
events. The relative difference between simulated and reconstructed
form factor values is shown in Fig.\ \ref{fpisim}b. There is no
systematic shift for all energy points except for the first three (0.61,
0.62 and 0.63 GeV).
\item For some energy points there is a small difference between energy
depositions of electrons and positrons, which comes mainly
from a difference in the energy calibration of different parts of the
calorimeter. The corresponding systematic error was estimated from the
simulation. 100000 of $e^+e^-\rightarrow e^+e^-$ events and
the corresponding number of $e^+e^-\rightarrow\mu^+\mu^-,\pi^+\pi^-$
events were simulated for 7 energy points.  The energy deposition of
particles was generated according to the energy deposition functions
described in the previous sections, but with different parameters
for $e^+$ and $e^-$. Then the event separation was
done by minimization of the same likelihood function as for the real
events. The systematic shift of about $0.5-0.8\%$ was observed
between simulated and reconstructed form factor values.
\end{enumerate}

We estimate the average systematic error of event separation to be
less than 0.6\%. For the first three energy points the systematic
error is larger (4\%, 2\% and 1\% for 0.61, 0.62 and 0.63 GeV energy
points respectively), but since the statistical error is a few times
larger we neglect this fact. 

\paragraph{Energy calibration of the collider.}

For almost all energy points the beam energy was measured using the resonance
depolarization technique \cite{depol}. 
The systematic error in beam energy measurement
does not exceed 50 keV. The corresponding systematic error in the pion
form factor does
not exceed 0.1\% for all energy points, except for the narrow \( \rho
-\omega  \) interference 
energy range, where it is estimated to be less than 0.5\% for 390 and 392
MeV energy points and less than 1\% for 391 MeV energy point.

\paragraph{Fiducial volume (accuracy of 
\protect\(\Theta_{min}\protect\) measurement).}  

\begin{figure}[h]
{\centering \resizebox*{0.8\textwidth}{!}{\includegraphics{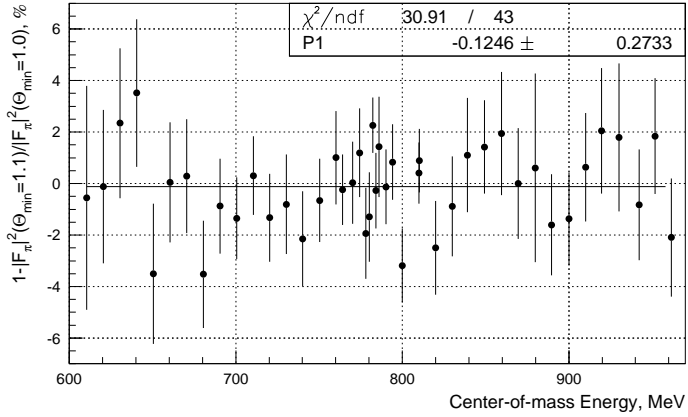}} \par}
\caption{\label{dfpith}The relative difference in the form factor
values, measured for $\Theta_{min}=1.0$ and $\Theta_{min}=1.1$
rad. The errors correspond to statistical fluctuation of the number of
events in $1.1<\Theta<1.0$ region. In other words, the errors show the
expected statistical difference for the form factor values measured
for $\Theta_{min}=1.0$ and $\Theta_{min}=1.1$ }
\end{figure}

The Z coordinate of the track intersection with the Z-chamber is
measured with better than 1 mm systematic accuracy. That allows, 
after the calibration of the drift chamber by the data from the
Z-chamber \cite{dccal}, to have systematic accuracy in polar angle
measurement better than 2.5 mrad. The corresponding systematic
error in the pion form factor is 0.5\%. Since the pion form factor was
measured independently for two \( \Theta _{min} \) values , we can
estimate this systematic error by the average form factor deviation \(
\left( 1-|F_{\pi }|^{2}\left| _{\Theta _{min}=1.0}\right. \left/
|F_{\pi }|^{2}\left| _{\Theta _{min}=1.1}\right. \right. \right)  \). 
This deviation is shown in Fig.\ \ref{dfpith}. It is equal on the average
to \( -0.1\%\pm 0.3\% \) and is consistent with the expected
statistical fluctuation. 

\paragraph{Trigger efficiency.}

For most energy points the trackfinder efficiency is high ($>98\%$), and as it
was discussed above, all types of collinear events are similar from
the trackfinder
point of view. Therefore it is assumed that the corresponding systematic error
is negligible. This statement is questionable for 420-435 and 470 MeV energy
points where the TF efficiency is lower than usual. The CsI efficiency is high for
all energy points ($>99\%$) and was measured for all types of collinear events.
Therefore the corresponding systematic error is negligible. The
combined systematic 
error is estimated to be less than 0.2\% for all energy points except 420-435
and 470 MeV, where it is estimated to be less than 1\%.

\paragraph{Reconstruction efficiency. }

The reconstruction efficiency was measured only for Bhabha events, but it was
found that it is uniform over \( \varphi  \) and \( \Theta  \) and
that it comes mostly from errors in the track recognition
algorithm. Therefore it is reasonable to assume that the
reconstruction efficiency is the same for all types of collinear
events. The efficiency drop for energy points below 800 MeV caused by
the lower precision of $\Theta$ angle
measurements is also the same for all types of collinear events. The
corresponding 
systematic error is estimated to be less than 0.3\%.

\paragraph{Correction for the nuclear interaction of pions.}

The estimated accuracy of the correction calculation is about 20\%, determined
by the cross-sections precision. The estimated contribution to this correction from
nuclear interaction of pions inside the drift chamber and from elastic pion
scattering by a large angle is about 30\% of the correction. Therefore
the corresponding systematic error is less than 0.4\%. 

\paragraph{Correction for the pions decays in flight.}

The systematic error of this correction is estimated to be less than 0.1\%.

\paragraph{Radiative corrections.}

The systematic error in radiative corrections is dominated by the
error in radiative corrections for \( e^{+}e^{-}\rightarrow e^{+}e^{-}
\), where it is estimated to be less than 1\%.

\section{A fit of the pion form factor}

%\afterpage{
%\clearpage
%\begin{longtable}{|c|c|c|c|c|c|c|}
%\hline 
%2E (MeV) & $N_{ee}$ & $N_{\mu\mu}$ & $N_{\pi\pi}$ & $N_{cosmic}$ & 
%$\sigma_{\pi\pi}$ (nb) & $|F_\pi|^2$ \\
%\hline 
%\endhead
%\multicolumn{7}{|r|}{{\it continued on the next page}} \\ \hline
%\multicolumn{7}{c}{ } \\
%\caption[]{The experimental data from the CMD-2 detector.}
%\endfoot
%\multicolumn{7}{c}{ } \\
%\caption{\label{data}The experimental data from the CMD-2 detector.}
%\endlastfoot
%\hline 
%\input{rho9495.tab}
%\end{longtable}
%}

\afterpage{
\clearpage
\begin{longtable}{|c|c|c|c|}
\hline 
2E (MeV) & $N_{\pi\pi}$ & $\sigma_{\pi\pi}$ (nb) & $|F_\pi|^2$ \\
\hline 
\endhead
\multicolumn{4}{|r|}{{\it continued on the next page}} \\ \hline
\multicolumn{4}{c}{ } \\
\caption[]{The experimental data from the CMD-2 detector}
\endfoot
\multicolumn{4}{c}{ } \\
\caption{\label{data}The experimental data from the CMD-2 detector}
\endlastfoot
\hline 
$610.50$ &
$  426\pm   50$ &
$333.6\pm 38.9$ &
$ 8.14\pm 0.95$
\\ \hline
$620.50$ &
$  829\pm   60$ &
$378.0\pm 26.8$ &
$ 9.41\pm 0.67$
\\ \hline
$630.50$ &
$  948\pm   64$ &
$422.8\pm 27.8$ &
$10.74\pm 0.71$
\\ \hline
$640.51$ &
$ 1005\pm   63$ &
$436.3\pm 26.6$ &
$11.31\pm 0.69$
\\ \hline
$650.49$ &
$  851\pm   55$ &
$492.2\pm 30.2$ &
$13.02\pm 0.80$
\\ \hline
$660.50$ &
$ 1288\pm   65$ &
$549.4\pm 26.2$ &
$14.83\pm 0.71$
\\ \hline
$670.50$ &
$ 1513\pm   74$ &
$602.6\pm 27.8$ &
$16.60\pm 0.77$
\\ \hline
$680.59$ &
$ 1654\pm   77$ &
$697.9\pm 30.0$ &
$19.63\pm 0.84$
\\ \hline
$690.43$ &
$ 2257\pm   87$ &
$717.8\pm 25.3$ &
$20.59\pm 0.73$
\\ \hline
$700.52$ &
$ 3411\pm  108$ &
$845.1\pm 23.9$ &
$24.76\pm 0.70$
\\ \hline
$710.47$ &
$ 4196\pm  131$ &
$946.0\pm 26.3$ &
$28.28\pm 0.79$
\\ \hline
$720.25$ &
$ 3340\pm  113$ &
$1018.3\pm 30.2$ &
$31.06\pm 0.92$
\\ \hline
$730.24$ &
$ 2764\pm  108$ &
$1111.9\pm 37.7$ &
$34.61\pm 1.17$
\\ \hline
$740.20$ &
$ 3138\pm  116$ &
$1181.5\pm 37.7$ &
$37.53\pm 1.20$
\\ \hline
$750.28$ &
$ 4290\pm  143$ &
$1315.9\pm 37.3$ &
$42.66\pm 1.21$
\\ \hline
$760.18$ &
$ 3745\pm  136$ &
$1337.8\pm 41.0$ &
$44.24\pm 1.36$
\\ \hline
$764.17$ &
$ 6241\pm  168$ &
$1301.8\pm 29.2$ &
$43.40\pm 0.97$
\\ \hline
$770.11$ &
$ 4805\pm  151$ &
$1301.1\pm 34.3$ &
$43.90\pm 1.16$
\\ \hline
$774.38$ &
$ 3729\pm  127$ &
$1250.5\pm 35.8$ &
$42.55\pm 1.22$
\\ \hline
$778.17$ &
$ 3897\pm  135$ &
$1313.6\pm 38.0$ &
$45.04\pm 1.30$
\\ \hline
$780.17$ &
$ 3993\pm  136$ &
$1253.9\pm 35.7$ &
$43.17\pm 1.23$
\\ \hline
$782.23$ &
$ 9709\pm  204$ &
$1080.7\pm 19.2$ &
$37.36\pm 0.66$
\\ \hline
$784.24$ &
$ 4946\pm  143$ &
$986.9\pm 24.3$ &
$34.25\pm 0.84$
\\ \hline
$786.04$ &
$ 2709\pm  108$ &
$869.6\pm 30.4$ &
$30.29\pm 1.06$
\\ \hline
$790.10$ &
$ 4948\pm  149$ &
$886.3\pm 23.3$ &
$31.12\pm 0.82$
\\ \hline
$794.14$ &
$ 4666\pm  145$ &
$836.5\pm 23.0$ &
$29.61\pm 0.81$
\\ \hline
$800.02$ &
$ 4701\pm  139$ &
$819.6\pm 21.2$ &
$29.35\pm 0.76$
\\ \hline
$810.14$ &
$ 6042\pm  152$ &
$699.2\pm 15.5$ &
$25.55\pm 0.57$
\\ \hline
$809.98$ &
$ 6648\pm  163$ &
$696.5\pm 15.0$ &
$25.44\pm 0.55$
\\ \hline
$820.02$ &
$ 2606\pm   99$ &
$648.2\pm 21.8$ &
$24.15\pm 0.81$
\\ \hline
$829.97$ &
$ 2209\pm   90$ &
$542.7\pm 19.9$ &
$20.61\pm 0.76$
\\ \hline
$839.10$ &
$ 1687\pm   77$ &
$450.3\pm 18.6$ &
$17.41\pm 0.72$
\\ \hline
$849.24$ &
$ 1194\pm   51$ &
$362.2\pm 14.5$ &
$14.28\pm 0.57$
\\ \hline
$859.60$ &
$ 1288\pm   63$ &
$364.3\pm 16.6$ &
$14.65\pm 0.67$
\\ \hline
$869.50$ &
$ 1581\pm   71$ &
$274.8\pm 11.6$ &
$11.26\pm 0.48$
\\ \hline
$879.84$ &
$  544\pm   42$ &
$260.6\pm 19.1$ &
$10.89\pm 0.80$
\\ \hline
$889.72$ &
$ 1610\pm   70$ &
$203.4\pm  8.4$ &
$ 8.66\pm 0.36$
\\ \hline
$900.04$ &
$ 1885\pm   76$ &
$182.0\pm  7.0$ &
$ 7.90\pm 0.30$
\\ \hline
$910.02$ &
$ 1352\pm   65$ &
$156.4\pm  7.3$ &
$ 6.92\pm 0.32$
\\ \hline
$919.56$ &
$ 1149\pm   60$ &
$135.6\pm  6.9$ &
$ 6.11\pm 0.31$
\\ \hline
$930.11$ &
$  748\pm   48$ &
$127.8\pm  8.0$ &
$ 5.87\pm 0.37$
\\ \hline
$942.19$ &
$ 1272\pm   61$ &
$116.1\pm  5.4$ &
$ 5.45\pm 0.26$
\\ \hline
$951.84$ &
$ 1220\pm   66$ &
$ 96.3\pm  5.1$ &
$ 4.60\pm 0.24$
\\ \hline
$961.52$ &
$ 1097\pm   60$ &
$ 91.6\pm  4.9$ &
$ 4.45\pm 0.24$
\\ \hline

\end{longtable}
}

The pion form factor data, obtained from the described analysis, is
summarized in Table \ref{data}. Only statistical errors are
shown. 

It is well known that to
describe the pion form factor the higher resonances $\rho(1450)$
and $\rho(1700)$ should be taken into account 
in addition to leading contribution from $\rho(770)$ and $\omega(782)$
\cite{OLYACMD,tau}. At the same time it was shown \cite{Newrho} that the
experimental data below 1 GeV is well described by the model based on
the hidden local symmetry, which predicts a point-like coupling
$\gamma\pi^+\pi^-$.
Here we use both approaches to fit the data. Since we fit
the pion form factor in the relatively narrow energy region 0.61-0.96 GeV,
only one higher resonance $\rho(1450)$ is taken into account.

\subsection{The Gounaris-Sakurai (GS) parametrization}

This model is based on the Gounaris-Sakurai parametrization of the
$\rho$-reso\-nance. Its modifications were used for parametrization of
all previous $e^+e^-$ data \cite{OLYACMD} and $\tau$ decay data \cite{tau}.
To describe the data, the $\rho(770)$, $\rho(1450)$ contributions and
$\rho-\omega$ interference were taken into account:
\begin{equation}
\label{GS}
F_\pi(s)=\frac{\GS_{\rho(770)}(s)\cdot
\displaystyle\frac{\mathstrut 1+\delta \, \BW_{\omega}(s)}{1+\delta}
+ \beta \, \GS_{\rho(1450)}(s)
}{1+\beta}.
\end{equation}

The $\rho(770)$ and $\rho(1450)$ contributions are taken according to
the Gouna\-ris-Sakurai model \cite{gunsac}:
\begin{equation}
\GS_{\rho(\mrho)} = \frac{
\mrho^2 \left( 1 + d \cdot \Gamma_\rho / \mrho \right)
}{
\mrho^2 - s + f(s) - i \mrho \Gamma_\rho(s)
},
\end{equation}
where 
\begin{equation}
f(s) = \Gamma_\rho \frac{\mrho^2}{p_\pi^3(\mrho^2)}
\left[
p_\pi^2(s) \left( h(s) - h(\mrho^2) \right) + 
(\mrho^2-s) \, p_\pi^2(\mrho^2) 
\left. \frac{dh}{ds} \right|_{s=\mrho^2}
\right] ,
\end{equation}
\begin{equation}
h(s) = \frac{2}{\pi} \frac{p_\pi(s)}{\sqrt{s}}
\ln \frac{\sqrt{s}+2 p_\pi(s)}{2m_\pi} ,
\end{equation}
\begin{equation}
\left. \frac{dh}{ds} \right|_{s=\mrho^2} = 
h(\mrho^2) \left[ \frac{1}{8p_\pi^2(\mrho^2)} - 
\frac{1}{2\mrho^2} \right] + \frac{1}{2\pi\mrho^2} ,
\end{equation}
\begin{equation}
p_\pi (s) = \frac{1}{2} \sqrt{s-4m_\pi^2}.
\end{equation}

The energy dependent width is
\begin{equation}
\label{rhowidth}
\Gamma_\rho(s) = \Gamma_\rho 
\left[ \frac{p_\pi(s)}{p_\pi(\mrho^2)} \right]^3
\left[ \frac{\mrho^2}{s} \right]^{1/2}.
\end{equation}

The normalization $\GS_{\rho(\mrho^2)}(0)=1$ fixes the parameter
$d$:
\begin{equation}
d = \frac{3}{\pi} \frac{m_\pi^2}{p_\pi^2(\mrho^2)}
\ln \frac{\mrho + 2 p_\pi(\mrho^2)}{2m_\pi}
+ \frac{\mrho}{2\pi p_\pi(\mrho^2)}
- \frac{m_\pi^2 \mrho}{\pi p_\pi^3(\mrho^2)}.
\end{equation}

The simple Breit-Wigner parametrization (\ref{BW}) is used for the
$\omega(782)$ contribution. 

In order to extract $\Gamma(\rho\rightarrow e^+ e^-)$ we relate the pion
form factor at the  $\rho$ mass to the VMD form factor:
\begin{equation}
\left. F_\pi(s) \right|_{s=\mrho^2} = 
\frac{ g_{\rho\gamma} g_{\rho\pi\pi} }{\mrho^2 - s - i \mrho
\Gamma_\rho} 
+ \left( \begin{array}{c} 
\mathrm{non-resonant} \\ \mathrm{contribution} 
\end{array} \right)
.
\end{equation}
From this one can obtain
\begin{equation}
g_{\rho\gamma} g_{\rho\pi\pi} = 
\frac{ \mrho^2 \left( 1 + d \cdot \Gamma_\rho / \mrho \right) }
{(1+\delta)(1+\beta)}.
\end{equation}
Using well known VMD relations \cite{Newrho}
\begin{equation}
\label{Gvee}
\Gamma_{V\rightarrow e^+e^-} = 
\frac{4\pi\alpha^2}{3\mathrm{M}_V^3} g_{V\gamma}^2,
\end{equation}
\begin{equation}
\label{Gvpp}
\Gamma_{V\rightarrow\pi^+\pi^-} = 
\frac{g_{V\pi\pi}^2}{6\pi} 
\frac{p_\pi^3 \left( \mathrm{M}_V^2 \right)}{\mathrm{M}_V^2}
\end{equation}
and assuming that $\Gamma_{\rho\rightarrow\pi^+\pi^-}=\Gamma_\rho$ one
obtains
\begin{equation}
\Gamma_{\rho\rightarrow e^+e^-} = 
\frac{2\alpha^2 p_\pi^3 \left( \mrho^2 \right)}{9 \mrho \Gamma_\rho}
\frac{(1+ d \cdot \Gamma_\rho / \mrho )^2}{(1+\delta)^2(1+\beta)^2}.
\end{equation}

A similar approach is used for the calculation of
$Br(\omega\rightarrow\pi^+\pi^-)$. One can relate the pion form factor at
the $\omega$ mass to the VMD form factor:
\begin{equation}
\left. F_\pi(s) \right|_{s=\momg^2} = 
\frac{ g_{\omega\gamma} g_{\omega\pi\pi} }{\momg^2 - s - i \momg
\Gamma_\omega}
+ \left( \begin{array}{c} 
\mathrm{non-resonant} \\ \mathrm{contribution} 
\end{array} \right)
.
\end{equation}
From this one determines $g_{\omega\pi\pi}$:
\begin{equation}
g_{\omega\gamma} g_{\omega\pi\pi} = 
\frac{ \delta \cdot \momg^2 \cdot |\GS_{\rho(770)}(\momg^2)| }
{(1+\delta)(1+\beta)}.
\end{equation}
Using (\ref{Gvee}) and (\ref{Gvpp}) one derives
\begin{equation}
Br(\omega\rightarrow\pi^+\pi^-) = 
\frac{2\alpha^2 p_\pi^3(\momg^2)}
{9\momg\Gamma_{\omega\rightarrow e^+e^-}\Gamma_\omega}
\left| \GS_{\rho(770)}(\momg^2) \right|^2
\frac{\delta^2}{(1+\delta)^2 (1+\beta)^2}.
\end{equation}

\subsection{The Hidden Local Symmetry (HLS) parametrization}

In the Hidden Local Symmetry (HLS) model \cite{Newrho,HLS,rhoomega}
the $\rho$-meson appears as a dynamical gauge boson of a hidden local 
symmetry in the non-linear chiral Lagrangian. This model introduces a
real parameter $a$ related to the non-resonant coupling $\gamma\pi^+\pi^-$.
The resulting pion form factor for the HLS model is 
\begin{equation}
\label{HLS}
F_\pi (s)= -\frac{a}{2}+1+\frac{a}{2} \cdot \BW_\rho(s) \cdot 
\frac{ 1 + \delta\,\BW_{\omega}(s)}{1+\delta},
\end{equation}
where
\begin{equation}
\label{BW}
\BW_{V}(s) =
\frac{\mathrm{M}_V^2}{\mathrm{M}_V^2-s-i\mathrm{M}_V\Gamma_V(s)}.
\end{equation}

The energy dependent width $\Gamma_\rho(s)$ is the same as for the GS
parametrization and is given by (\ref{rhowidth}).

Using the similar approach as for the GS parametrization, we obtain
\begin{equation}
\Gamma_{\rho\rightarrow e^+e^-} = 
\frac{2\alpha^2 p_\pi^3 \left( \mrho^2 \right)}{9 \mrho \Gamma_\rho}
\left[ \frac{a}{2(1+\delta)} \right]^2,
\end{equation}
\begin{equation}
Br(\omega\rightarrow\pi^+\pi^-) = 
\frac{2\alpha^2 p_\pi^3(\momg^2)}
{9\momg\Gamma_{\omega\rightarrow e^+e^-}\Gamma_\omega}
\left| \frac{a}{2}\BW_\rho(\momg^2) \right|^2
\frac{\delta^2}{(1+\delta)^2}.
\end{equation}

\subsection{Result of fit}

The fit to data was done by the minimization of the following $\chi^2$
function:
\begin{equation}
\chi^2 = \sum_{i=1\ldots 44} 
\frac{ \left( |F_\pi|^2_{exp}(s_i) - |F_\pi|^2_{theor}(s_i) \right)^2 }
{ \Delta_i^2 } + \sum_{j=1\ldots N_p} 
\frac{ \left( p_j - p_{0j} \right)^2 }{\sigma_j^2},
\end{equation}
where $|F_\pi|^2_{exp}(s_i)$ and $|F_\pi|^2_{theor}(s_i)$ are the
experimental and theoretical values of the pion form factor at the $i$-th
energy point, $\Delta_i$ is the experimental error at the $i$-th energy
point, $N_p$ is the number of model parameters, $p_j$ is the $j$-th
model parameter, $p_{0j}$ is the expected value of $j$-th model
parameter and $\sigma_j$ is the estimated error of $j$-th model
parameter. The model parameters are the following (all values are
taken from PDG'98 \cite{PDG98}).
\begin{enumerate}
\item The $\omega$ meson mass: $p_{01}=781.94$ MeV, $\sigma_1=0.12$ MeV.
\item The $\omega$ meson width: $p_{02}=8.41$ MeV, $\sigma_2=0.09$ MeV.
\item The $\omega$ meson leptonic width: $p_{03}=0.60$ keV,
$\sigma_3=0.02$ keV. 
\item The $\rho(1450)$ meson mass: $p_{04}=1465$ MeV, $\sigma_4=25$ MeV.
\item The $\rho(1450)$ meson width: $p_{05}=310$ MeV, $\sigma_5=60$ MeV.
\end{enumerate}
For the GS fit the number of model parameters is $N_p=5$, while for
the HLS fit $N_p=3$.

The parameter $\beta$ in (\ref{GS}) was assumed to be real.
To determine a phase of $\rho-\omega$ mixing, the parameter $\delta$ in
(\ref{GS}) and (\ref{HLS}) was assumed to be complex during the
fit. However, it turns out that the phase was always consistent with
zero. Therefore $\delta$ was assumed to be real for the final fit. 

\begin{figure}
{\centering \resizebox*{\textwidth}{!}{\includegraphics{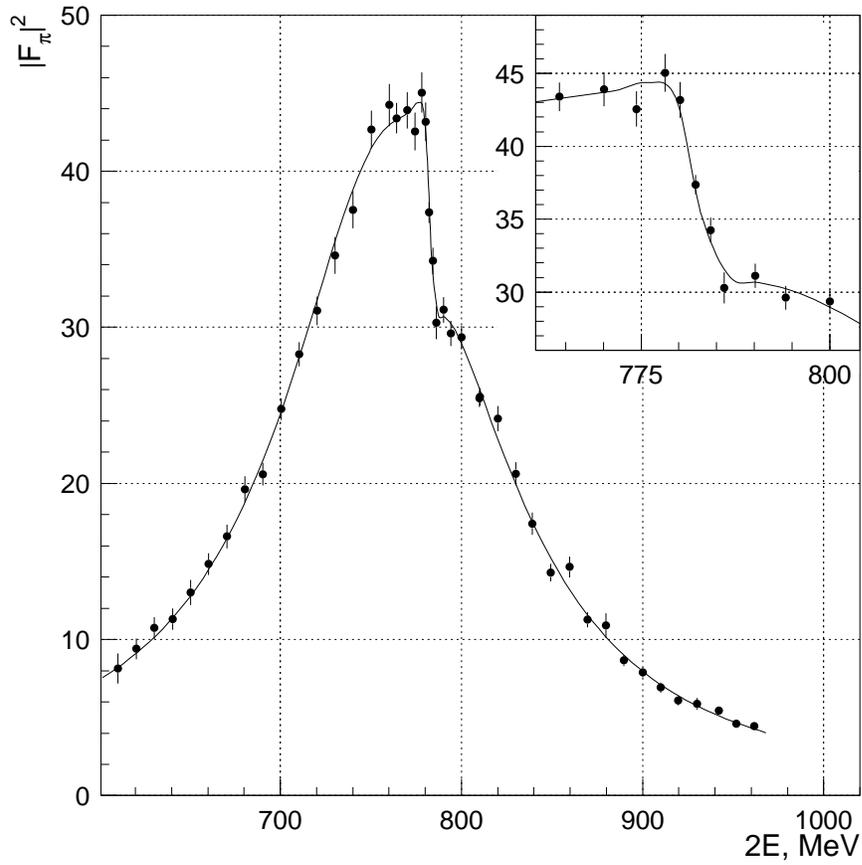}} \par}
\caption{\label{fpifit}Fit of the CMD-2 (94, 95) pion form factor data
according to GS and HLS models. Both theoretical curves are presented,
but they are indistinguishable}
\end{figure}

The results of the fit of the CMD-2 experimental data are shown in
Fig.\ \ref{fpifit} and are summarized in Table \ref{rhopar}.
The first error is statistical, the second error is systematic. In order
to estimate a systematic error, the fit was repeated with all data
points shifted up and down by one systematic error.

\begin{table}
\begin{center}
\begin{tabular}{|l|c|c|}
\hline
& GS model & HLS model \\ \hline
$\mrho$, MeV & $775.28\pm 0.61\pm 0.20$ & $774.57\pm 0.60\pm 0.20$ \\ \hline
$\Gamma_\rho$, MeV & $147.70\pm 1.29 \pm 0.40$ & $147.65\pm 1.38\pm 0.20$ \\ \hline
$Br(\omega\rightarrow\pi^+\pi^-)$, \% & $1.31\pm 0.23\pm 0.02$ &
$1.32\pm 0.23\pm 0.02$ \\ \hline
$\Gamma(\rho\rightarrow e^+e^-)$, keV & $6.93\pm 0.11\pm 0.10$ &
$6.89\pm 0.12\pm 0.10$ \\ \hline
$\beta$ (GS) & $-0.0849\pm 0.0053\pm 0.0050$ & --- \\ \hline
$a$ (HLS) & --- & $2.381\pm 0.016\pm 0.016$ \\ \hline
$\chi^2/n$ & 0.77 & 0.78 \\ \hline
\end{tabular}
\end{center}
\caption{\label{rhopar}The results of fit of the CMD-2 (94,95) pion
form factor data by
GS and HLS models}
\end{table}

\subsection{Hadronic Contribution to (g-2)$_{\mu}$}

We'll now estimate the implication of our results for the value
and uncertainty of $a_{\mu}^{had}$ - the hadronic contribution
to (g-2)$_{\mu}$. To this end we'll choose the c.m.energy 
range near the $\rho$-meson peak (from 630 to 810 MeV) which gives the 
dominant contribution to the muon anomaly and where good data are
available from previous 
measurements by the OLYA and CMD groups \cite{OLYACMD} as well as 
by the DM1 detector \cite{DM1}. Table \ref{g-2} presents results
of our calculations of $a_{\mu}^{had}$ performed by direct integration 
of the experimental data over the energy range above. The calculation
used the following procedure: first we integrated the data of each group
over the chosen energy range. Since measurements at different energy
points are independent, the squared statistical error of the integral 
can be obtained by summing statistical errors squared of each energy 
point. After that the systematic uncertainty which is believed to be
an overall normalization uncertainty is added quadratically. 
Contributions of separate groups are subject to weighted averaging 
taking into account if necessary a scale factor \cite{PDG98}.    
Such a procedure implies an assumption that systematical uncertainties
are uncorrelated since they refer to different measurements.
  
\begin{table}[h]
\begin{center}
\begin{tabular}{|l|c|c|}
\hline
Data  & a$_{\mu}^{had}$, 10$^{-10}$ & Total error, 10$^{-10}$ \\
\hline
Old   & 284.2 $\pm$ 3.8 $\pm$ 7.1 & 8.1 \\
\hline
New   & 292.5 $\pm$ 2.2 $\pm$ 4.1 & 4.7 \\
\hline
Old+New & 290.3 $\pm$ 1.9 $\pm$ 3.6 & 4.1 \\
\hline
\end{tabular}
\end{center}
\caption{\label{g-2}Hadronic contributions to (g-2)$_{\mu}$, coming
from the c.m.energy range near the $\rho$-meson peak.}
\end{table}    

The first line gives the average of three independent estimates based 
on the data of OLYA, CMD and MD1 while the second one presents the 
corresponding value for the CMD-2 data. The third line of the Table is 
the average based on the four independent estimates above. For 
convenience, we list separately the statistical and systematic uncertainties
in the second column while the third one gives the total error obtained by 
adding them quadratically. One can see that the estimate based on the
CMD-2 data is in good agreement with that coming from the old data.  
Note also the significant improvement of both statistical and systematic 
uncertainties compared to the previous measurements. The combined 
result (old plus new data) is a factor of two more precise than before.

\section{Conclusion}

A new measurement of the pion form factor in the center-of-mass
energy range $0.61-0.96$ GeV is presented. The statistical error is
about the same as for all previous $e^+e^-$ data, but the systematic
error is a few times smaller. The results are based only
on part of the experimental data taken by CMD-2. There are several
reasons why this data set was treated as independent. First, a
different approach in data analysis is required for the data taken
in the center-of-mass energy range below $0.6$ GeV and above $1.0$
GeV (runs 4 and 5 in table \ref{runstable}). Second, the data taking
conditions were significantly different in 1998 (run 6 in table
\ref{runstable}), when  the data was taken in the same $0.61-0.96$ GeV
energy range. As a result, the systematic errors for these runs will
be different than for the data presented here. Therefore we prefer 
to present separate results for different groups of runs. 

The fit of the pion form factor based on the Gounaris-Sakurai and the
Hidden Local Symmetry parametrizations was performed. For the final
values we prefer the Gounaris-Sakurai parametrization as it was
traditionally used for the determination of the $\rho$-meson
parameters 
(see \cite{OLYACMD} and \cite{tau}). 
The difference between the results of the fit in two models
is always smaller than the experimental error (Table
\ref{rhopar}), therefore the ``model'' error is not specified. We give
as the final results the following, where the first error is
statistical and the second one is systematic:
\begin{equation}
\left\{
\begin{array}{ll}
\mrho \, (\mathrm{MeV}) & = 775.28\pm 0.61\pm 0.20 ,
\vphantom{\Bigl( \Bigr)} \\
\Gamma_\rho \, (\mathrm{MeV}) & = 147.70\pm 1.29 \pm 0.40 ,
\vphantom{\Bigl( \Bigr)} \\
\Gamma(\rho\rightarrow e^+e^-) \, (\mathrm{keV}) & = 6.93\pm 0.11\pm 0.10 ,
\vphantom{\Bigl( \Bigr)} \\
Br(\omega\rightarrow\pi^+\pi^-) & = ( 1.31\pm 0.23 ) \% .
\vphantom{\Bigl( \Bigr)} 
\end{array}
\right.
\end{equation}

It has also been shown that the improvement of the experimental
precision, particularly of the systematic uncertainties, can be
crucial for the high precision calculation of the hadronic contribution 
to the muon anomaly.

With the progress of analysis of the rest available data we
hope to reduce the systematic error for the data presented here by
additional factor of two. Important part of this improvement is the
development of the new approach to the radiative corrections
calculation with the ultimate goal to reduce the corresponding
systematic error to $(0.3-0.5)\%$ level. Using the high statistics
taken in 1998, it will be possible to perform the detailed analysis of
small systematic effects. 

The authors are grateful to the staff of VEPP-2M for excellent
performance of the collider, to all engineers and technicians who
participated in the design, commissioning and operation of CMD-2. We
acknowledge the contribution to the experiment at the earlier stages
of our colleagues V.A.Monich, A.E.Sher, V.G.Zavarzin and
W.A.Worstell. Special thanks are due to N.N.Achasov, A.B.Arbuzov,
M.Benayoun, V.L.Chernyak, V.S.Fadin, F.Jegerlehner,
E.A.Kuraev, A.I.Milstein and G.N.Shestakov
for useful discussions and permanent interest.

This work is supported in part by grants RFBR-98-02-17851, INTAS
96-0624 and DOE DEFG0291ER40646.

\end{document}